\documentclass[11pt]{article}

\usepackage{stackengine}

\usepackage{stackrel}
\stackMath
\makeatletter
\normalsize
\divide\@tempdima\baselineskip
\@tempcnta=\@tempdima
\skip\@mpfootins = \skip\footins
\renewcommand\section{\@startsection{section}{1}{\z@}%
                                   {-3.5ex \@plus -1.3ex \@minus -.7ex}%
                                   {2.3ex \@plus.4ex \@minus .4ex}%
                                   {\normalfont\large\bfseries}}
\renewcommand\subsection{\@startsection{subsection}{2}{\z@}%
                                   {-2.3ex\@plus -1ex \@minus -.5ex}%
                                   {1.2ex \@plus .3ex \@minus .3ex}%
                                   {\normalfont\normalsize\bfseries}}
\renewcommand\subsubsection{\@startsection{subsubsection}{3}{\z@}%
                                   {-2.3ex\@plus -1ex \@minus -.5ex}%
                                   {1ex \@plus .2ex \@minus .2ex}%
                                   {\normalfont\normalsize\bfseries}}
\renewcommand\paragraph{\@startsection{paragraph}{4}{\z@}%
                                   {1.75ex \@plus1ex \@minus.2ex}%
                                   {-1em}%
                                   {\normalfont\normalsize\bfseries}}
\renewcommand\subparagraph{\@startsection{subparagraph}{5}{\z@}%
                                   {1.75ex \@plus1ex \@minus .2ex}%
                                   {-1em}%
                                   {\normalfont\normalsize\itshape}}

\renewcommand{\@dotsep}{10000}

\def\fnum@figure{\textbf{\figurename\nobreakspace\thefigure}}
\def\fnum@table{\textbf{\tablename\nobreakspace\thetable}}

\long\def\@makecaption#1#2{%
  \vskip\abovecaptionskip
  \sbox\@tempboxa{\small #1. #2}%
  \ifdim \wd\@tempboxa >\hsize
    \small #1. #2\par
  \else
    \global \@minipagefalse
    \hb@xt@\hsize{\hfil\box\@tempboxa\hfil}%
  \fi
  \vskip\belowcaptionskip}

\renewenvironment{thebibliography}[1]{%
\begin{oldthebibliography}{#1}%
\small%
\raggedright%
\setlength{\itemsep}{5pt plus 0.2ex minus 0.05ex}%
}%
{%
\end{oldthebibliography}%
}
\makeatother

\usepackage[colorlinks=true, linkcolor=blue, filecolor=blue, urlcolor=blue, citecolor=blue, anchorcolor=blue, menucolor=blue , linktocpage=true]{hyperref}
\usepackage{scalerel}
\usepackage{graphicx}
\usepackage{amsmath,amssymb}
\usepackage{amsfonts}
\usepackage{physics}
\usepackage{bbold}
\usepackage{mathtools}
\usepackage{verbatim}
\usepackage{cite} 
\usepackage{mdframed}
\usepackage{enumitem}
\usepackage[normalem]{ulem}

\allowdisplaybreaks
\numberwithin{equation}{section}

\makeatletter
\newcommand{\subalign}[1]{%
  \vcenter{%
    \Let@ \restore@math@cr \default@tag
    \baselineskip\fontdimen10 \scriptfont\tw@
    \advance\baselineskip\fontdimen12 \scriptfont\tw@
    \lineskip\thr@@\fontdimen8 \scriptfont\thr@@
    \lineskiplimit\lineskip
    \ialign{\hfil$\m@th\scriptstyle##$&$\m@th\scriptstyle{}##$\hfil\crcr
      #1\crcr
    }%
  }%
}
\makeatother

\usepackage[textsize=scriptsize,textwidth=2.45cm]{todonotes}

\renewcommand{\title}[1]{\vbox{\center\LARGE{#1}}\vspace{5mm}}
\renewcommand{\author}[1]{\vbox{\center#1}\vspace{5mm}}

\newcommand{\email}[1]{\vspace{5mm}\vbox{\center\footnotesize\tt#1}\vspace{5mm}}

\newcommand{\nn}{\nonumber} %PETER!!

\begin{document}

\begin{titlepage}

\phantom{a}
\vskip20mm

\begin{center} 

\title{Revisiting Extremal Couplings in AdS/CFT}

%\vskip10mm

\author{Alejandra Castro$^{\,a}$ and
Pedro J. Martinez$^{\,b}$}

%\vskip1em

\vskip4.1mm

\begin{minipage}[c]{0.89\textwidth}\centering \footnotesize \em 
{\it $^{a}$Department of Applied Mathematics and Theoretical Physics, University of Cambridge,
Cambridge CB3 0WA, United Kingdom}\\ \vspace{0.5em}
{\it $^{b}$ 
Instituto de Física La Plata - CONICET, Universidad Nacional de La Plata, \\
La Plata, C.C. 67, 1900, Argentina\\}
 \end{minipage}

\email{ac2553@cam.ac.uk, martinezp@fisica.unlp.edu.ar}
\end{center}

\vskip6mm

\begin{abstract}
We consider an effective theory of massive scalar fields on a fixed AdS$_{d+1}$ background with a cubic \emph{extremal} interaction among them. A bulk coupling is called extremal whenever the corresponding conformal dimension of any of the dual CFT$_d$ operators matches the sum of all the others. For cubic bulk couplings, this is $\Delta_i+\Delta_j=\Delta_k$. These bulk interactions are often disregarded in the literature since they do not appear in traditional models of AdS/CFT. Turning them on yields a divergent vertex in the dual CFT, and here we show that these divergences can be regulated. Once renormalized, we demonstrate that this coupling introduces non-trivial mixing between single- and double-trace operators, and we compute the anomalous dimensions of the corrected operators to leading order in perturbation theory.
\end{abstract}

\end{titlepage}

\eject

\phantom{a}
\vspace{-4em}
{
\tableofcontents
}

\bigskip

\bigskip

%\newpage

%%%%%%%%%%%%%%%%%%%%%%%%%%%%%%%%%%%%%%%%%%%%%%%%%%%%%%%%%%%%%%%%%%%%%%%%%%%%%%%%%%%%%%%%
%%%%%%%%%%%%%%%%%%%%%%%%%%%%%%%%%%%%%%%%%%%%%%%%%%%%%%%%%%%%%%%%%%%%%%%%%%%%%%%%%%%%%%%%
%%%%%%%%%%%%%%%%%%%%%%%%%%%%%%%%%%%%%%%%%%%%%%%%%%%%%%%%%%%%%%%%%%%%%%%%%%%%%%%%%%%%%%%%
\section{Introduction}
%%%%%%%%%%%%%%%%%%%%%%%%%%%%%%%%%%%%%%%%%%%%%%%%%%%%%%%%%%%%%%%%%%%%%%%%%%%%%%%%%%%%%%%%
%%%%%%%%%%%%%%%%%%%%%%%%%%%%%%%%%%%%%%%%%%%%%%%%%%%%%%%%%%%%%%%%%%%%%%%%%%%%%%%%%%%%%%%%
%%%%%%%%%%%%%%%%%%%%%%%%%%%%%%%%%%%%%%%%%%%%%%%%%%%%%%%%%%%%%%%%%%%%%%%%%%%%%%%%%%%%%%%%

The AdS$_{d+1}$/CFT$_{d}$ correspondence is by now a well-established tool to explore perturbative and non-perturbative properties of both conformal field theory (CFT$_{d}$) and quantum gravity in Anti-de Sitter (AdS$_{d+1}$).
Over its 25 years of development, this strong-weak correspondence has been realized and utilized in several ways.
The first entries of the holographic dictionary were drawn from specific constructions in string theory. The most prominent example is the relation between type IIB string theory compactified on AdS$_5\times S^5$ and ${\cal N} = 4$ super Yang-Mills theory \cite{Maldacena:1997re}, and several of the early days examples are reviewed in \cite{Aharony:1999ti}.  
With growing numbers of realizations of the AdS/CFT correspondence, it was soon clear that the physics of asymptotically AdS backgrounds coupled to matter could also be used as an effective description of CFTs regardless of a specific UV completion in string theory; see for example \cite{Penedones:2016voo}. This approach is often dubbed ``bottom-up,'' since one uses the correspondence to explore its possible implications, as opposed to the ``top-down'' realizations of the holographic dictionary in which the relation is not assumed but rather tested. 

The natural observables that establish the correspondence are correlation functions of local operators. 
From the CFT$_d$ side, the theory is locally defined in terms of the conformal dimension of the primary operators $\Delta_i$ and the OPE coefficients $c_{ijk}$ between these, appearing as the coefficients of the three-point functions between primaries. 
From the AdS$_{d+1}$ perspective, at leading order in the coupling, the OPE coefficients of the operators dual to local fields in AdS usually arise from cubic interactions of these bulk fields. If the cubic interaction has a bulk coupling $\lambda_{ijk}$, then the dictionary gives
\begin{equation}\label{eq:OPE-Ads}
    c_{ijk} = \lambda_{ijk} \, {\cal C}_{ijk} + \ldots~,
\end{equation}
where ${\cal C}_{ijk}$ is the coefficient that arises from the AdS$_{d+1}$ vertex integral \cite{Witten:1998qj,Freedman:1998tz} and depends on the conformal dimensions $\Delta_i$ and dimension $d$. The dots here denote terms suppressed by the so-called large-$N$ limit of the CFT$_d$; in gravitational terms, this is the tree-level contribution at $G_N\to 0$. There are also cases where a matching along the lines of \eqref{eq:OPE-Ads} is due to boundary couplings in AdS, which we will mention below.  

The coefficients ${\cal C}_{ijk}$, which we review in Sec.\,\ref{sec:prelim}, are meromorphic functions of its parameters. In particular, one of its most prominent poles occurs at {\it extremality}, i.e., when
\begin{equation}
    \Delta_i + \Delta_j = \Delta_k~.
\end{equation}
This indicates a divergence in the AdS$_{d+1}$ vertex. Our aim is to show how this divergence is renormalized and the interpretation of the regularized vertex in the CFT$_d$. 

Our approach to tame this divergence is bottom-up.  
The reason is that for all known top-down supersymmetric realizations of AdS/CFT known to date, one has that cubic interaction among fields in the appropriate supergravity regime obey\footnote{Actually the couplings $\lambda_{ijk}$ vanish whenever $\Delta_i + \Delta_j \leq \Delta_k$ for the known examples referenced below.}
\begin{equation} \label{eq:lambda-vanish}
    \lambda_{\rm ext}\equiv \lambda_{ijk} =0 ~, \qquad {\rm if} \quad \Delta_i + \Delta_j = \Delta_k~.
\end{equation}
In this case, the contribution to the three-point function comes from a boundary coupling in AdS$_{d+1}$, and it was extensively studied during the development of AdS/CFT. In particular, extremal correlators in AdS$_5\times S^5$ supergravity are discussed in \cite{Lee:1998bxa,DHoker:1998ecp,Liu:1999kg,DHoker:1999jke}, and for AdS$_3\times S^3$ in \cite{Mihailescu:1999cj,Arutyunov:2000by}. The fact that the interaction is located at the boundary of AdS$_{d+1}$ also leads to interesting aspects of the dictionary: field re-definitions play an important role \cite{Arutyunov:2000ima}, which is related to mixing among single- and multi-trace operators in the CFT$_d$ that alter the OPE coefficient \cite{Taylor:2007hs,Rastelli:2017udc,Rawash:2021pik}, see also \cite{Aprile:2018efk,Aprile:2020uxk}. It has also led to conjectures regarding the vanishing of couplings in supergravity that lead to extremal correlators even for $n$-point interactions \cite{DHoker:2000pvz,DHoker:2000xhf}. This conjecture has received non-trivial support recently, where tools in exceptional field theory are used to establish that these couplings indeed vanish for instances of AdS vacua arising from string theory \cite{Duboeuf:2023cth}.

The fact that extremal correlators arise from boundary interactions is key for the success of AdS/CFT from a top-down perspective, which motivates the general conjecture that \eqref{eq:lambda-vanish} should always hold. However, the evidence of cancellations is tied to the fact that the effective theories in AdS arise from supersymmetric compactifications in 10/11D supergravity, so we do not find a compelling reason to remove these bulk couplings from a bottom-up approach to AdS/CFT. Moreover, in \cite{Castro:2021wzn} an extremal bulk interaction was identified in a non-supersymmetric vacua: when AdS$_2\times S^2$ is embedded in the non-BPS branch of ${\cal N}=2$ 4D ungauged supergravity. The cubic interaction involved a massless scalar field in AdS$_2$ and the dilaton field (which is part of JT gravity in AdS$_2$). Although JT gravity is not a traditional instance of AdS/CFT, the appearance of this interaction motivated us to revisit what would happen if $\lambda_{\rm ext}\neq 0$ when $\Delta_i+\Delta_j=\Delta_k$ from a bottom-up perspective. 

The bottom-up setup we will use is the following. We will be considering massive scalar fields in AdS$_{d+1}$ that interact via a simple cubic interaction, and we will arrange the masses of the fields such that we have an extremal coupling in the bulk, i.e., $\lambda_{\rm ext}\neq0$. We will show that the vertex can be regularized using an asymptotic bulk cutoff description in the main body, and via a finite bulk cutoff description in an appendix. We obtain the same results for both methods. Our main emphasis will be to show how to renormalize the on-shell action by constructing appropriate counterterms that follow standard procedures in AdS/CFT  \cite{Skenderis:2002wp}; see also \cite{vanRees:2011fr,vanRees:2011ir,Bzowski:2015pba} which share some aspects with the analysis here.

The outcome of holographic renormalization is that the generating functional now contains a logarithmic three-point function at a finite distance in configuration space for the dual operators. We show that this corresponds to a non-trivial mixing between operators of the free theory. The reason is the following. At zero coupling  ($\lambda_{\rm ext}=0$) we have a degenerate spectrum where the single trace operator $\Delta_k$ can mix with the double-trace operator which has $\Delta_i+\Delta_j (=\Delta_k)$. Turning on $\lambda_{\rm ext}$ lifts this degeneracy and the logarithmic term captures the anomalous dimension of the new primary eigenstates of the system at finite coupling.   

After treating the extremal coupling, as a corollary, we will also discuss other peculiar cubic couplings in AdS$_{d+1}$ that are motivated by the poles in ${\cal C}_{ijk}$. These correspond to the three-point interactions where the fields involved have  
\begin{equation}\label{eq:1}
    \Delta_k=\Delta_i +\Delta_j + 2 n ~, \qquad n\in \mathbb{N}~,
\end{equation}
or 
\begin{equation}\label{eq:2}
    \Delta_i +\Delta_j + \Delta_k=d -2n ~, \qquad n\in \mathbb{N}^0~.
\end{equation}
These types of interactions also lead to pathologies in the AdS$_{d+1}$ vertex. We find that the nature and interpretation of \eqref{eq:1} are very similar to the extremal case. Technically \eqref{eq:2} shares many similarities to the extremal analysis but the physical interpretation is delicate, which we discuss. Shadow extremal correlators have been discussed in, for example, \cite{Apolo:2024hoa,Freedman:2016yue}. A bottom-up approach of shadow-extremal couplings is considered in \cite{Aharony:2015afa}, which is relevant to our discussion. 

The outline of this paper is as follows. In Sec.\,\ref{sec:prelim} we review some standard results from AdS/CFT and set the notation for the rest of the paper. In Sec.\,\ref{AdSd+1} we use a rigid cutoff prescription to regulate the divergent integral that appears when the extremal condition is met. The regularized answer leads to logarithmic contributions to the two- and three-point functions of the boundary CFT. We interpret this result as the first order in a $\lambda_{\rm ext}\to0$ expansion of a CFT were single and multi-trace operators mix and acquire an anomalous dimension.
In Sec.\,\ref{sec:peculiar} we complete our study of the divergencies present in the standard bulk vertex integral and find that an explanation analogous to the one found for the extremal case is always possible.
We close our work with a discussion in Sec.\,\ref{sec:Discussion}.
In App.\,\ref{App:Examples} we provide an independent check of our results in a more controlled set-up by working in a regularized bulk with exact Green functions. We choose specific examples in which all integrals can be carried analytically and find perfect match with our results in the main body of the paper. In App.\,\ref{app:conventions} we include a list of many mathematical identities used throughout our work. 

%%%%%%%%%%%%%%%%%%%%%%%%%%%%%%%%%%%%%%%%%%%%%%%%%%%%%%%%%%%%%%%%%%%%%%%%%%%%%%%%%%%%%%%%
%%%%%%%%%%%%%%%%%%%%%%%%%%%%%%%%%%%%%%%%%%%%%%%%%%%%%%%%%%%%%%%%%%%%%%%%%%%%%%%%%%%%%%%%
%%%%%%%%%%%%%%%%%%%%%%%%%%%%%%%%%%%%%%%%%%%%%%%%%%%%%%%%%%%%%%%%%%%%%%%%%%%%%%%%%%%%%%%%
\section{Preliminaries}\label{sec:prelim}
%%%%%%%%%%%%%%%%%%%%%%%%%%%%%%%%%%%%%%%%%%%%%%%%%%%%%%%%%%%%%%%%%%%%%%%%%%%%%%%%%%%%%%%%
%%%%%%%%%%%%%%%%%%%%%%%%%%%%%%%%%%%%%%%%%%%%%%%%%%%%%%%%%%%%%%%%%%%%%%%%%%%%%%%%%%%%%%%%
%%%%%%%%%%%%%%%%%%%%%%%%%%%%%%%%%%%%%%%%%%%%%%%%%%%%%%%%%%%%%%%%%%%%%%%%%%%%%%%%%%%%%%%%

This section will introduce the basic concepts and set our conventions, which will be used throughout. We will be following mainly \cite{Freedman:1998tz,Mueck:1999nr}. Our starting point is to consider a collection of massive scalar fields $\Phi_i(x)$, propagating on a fixed AdS$_{d+1}$ background. We will be working on Euclidean Poincar\'e AdS$_{d+1}$ and parameterize the metric as
\begin{equation}\label{metric-d+1}
\dd s^2=g_{\mu\nu}\dd x^\mu \dd x^\nu =\frac{\dd z_0^2+\dd \vec z\,^2}{z_0^2}~,
\end{equation}
where $z_0$ is the radial coordinate and the boundary of AdS$_{d+1}$ is located at $z_0\to 0$, and $\vec z$ are the boundary coordinates. The AdS radius will be set to unity throughout this work. We will denote the induced metric at fixed $z_0$ as $\gamma_{ab}$ ($a,b=1,\ldots, d$), and in Poincar\'e coordinates it reads $\gamma_{ab}=z_0^{-2} \delta_{ab}$.  
The Euclidean action for each of the scalar fields is
\begin{equation}\label{eq:S_i}
S_i=\frac 12 \int \dd^{d+1}x \,\sqrt{g}\, \left( \partial_\mu \Phi_i \partial^\mu \Phi_i + \Delta_i(\Delta_i-d) \Phi_i^2\right)~.
\end{equation}
Here we are trading the mass of the field with $\Delta_i$ via the standard relation, $m^2_i=\Delta_i(\Delta_i-d)$, where $\Delta_i\geq d/2$ is the largest root of this equation. In the following, we will quantize the field in the so-called ``standard quantization,'' where it will be interpreted as an operator in the dual CFT with conformal dimension $\Delta_i$. Quantizing the field such that the dual operator has conformal dimension $ \Delta^{ s}_i\equiv d-\Delta_i\leq d/2$ will be called ``alternative quantization.'' 

The bulk-to-boundary propagator for each field is given by 
\begin{equation}\label{eq:bndy-bulk-prop}
K_{\Delta_i}(z_0, \vec z ,\vec{x}) = 
\frac{\Gamma(\Delta_i)}{\pi^{\frac{d}{2}} \Gamma(\Delta_i - 
\frac{d}{2})} 
\left( \frac{z_0}{z_0^2 + (\vec{z}-\vec{x})^2} \right)^{\Delta_i} \;.
\end{equation}
In the absence of interactions, each field satisfies a Klein-Gordon equation and we will consider the solution to be  
\begin{equation}\label{d+1Field-Asymp}
\begin{aligned}
\Phi_i(z_0, \vec{z}) &= \int \dd^{d} x\; K_{\Delta_i}(z_0, \vec z ,\vec{x}) \phi(\vec{x}) \\
&=\frac{\Gamma(\Delta_i)}{\pi^{\frac{d}{2}} \Gamma(\Delta_i - \frac{d}{2})} \int \dd^{d} x\left( \frac{z_0}{z_0^2 + (\vec{z}-\vec{x})^2} 
\right)^{\Delta_i} \phi_i(\vec{x})~,
\end{aligned}
\end{equation}
which has the asymptotic boundary condition 
\begin{equation}\label{eq:leading-Phi}
    \Phi_i(z_0, \vec{z}) \sim z_0^{d-\Delta_i} \phi_i(\vec{z})+\dots~,\qquad z_0 \to 0~.
\end{equation}
We will denote the dual operator to each scalar field $\Phi_i(z_0,\vec z)$ as ${\cal O}_i(\vec z)$, and $\phi_i(\vec z)$ is the associated source. 
Using this notation, we write the holographic correspondence as
\begin{equation}\label{adscft}
    {\cal Z}_{\rm CFT}^d = \langle e^{\int \dd^d x \, \phi_i(\vec x) {\cal O}_i(\vec x)}\rangle \equiv \int_{\phi_i} {\cal D}\Phi_i \; e^{-S_{i}}= {\cal Z}_{\rm AdS}^{d+1}~,
\end{equation}
where on the right-hand side the subscript $\phi_i$ indicates that the path integral over the fields $\Phi_i$ is performed with the asymptotic boundary conditions \eqref{eq:leading-Phi} which holds $\phi_i$ fixed.  We have omitted the gravitational degrees of freedom to shorten the notation as they will not play an important role in this work. The quantities on both sides of \eqref{adscft} have to be regulated before giving them a proper physics interpretation. We will use $I_{\rm ren}=I+I_{\rm ct}$ to refer to a renormalized on-shell action, where $I$ is a bare on-shell action and $I_{\rm ct}$ the counterterms that render $I_{\rm ren}$ finite. Using a saddle point approximation we will therefore have 
\begin{equation}
    {\cal Z}_{\rm AdS}^{d+1}=\int_{\phi_i} {\cal D}\Phi_i \; e^{-S_{i}}\sim e^{-I_{\rm ren}}\;.
\end{equation}
To leading order, the resulting two-point function for the dual operators ${\cal O}_i(\vec x)$ is
\begin{equation}\label{2pf}
\langle {\cal O}_i(\vec{x}){\cal O}_j(\vec{y})\rangle \equiv-\frac{\delta I_{\rm ren}^{(0)}}{\delta \phi_i(\vec{x})\delta \phi_j(\vec{y})}=  \frac{(2\Delta_i-d)\Gamma(\Delta_i)}{\pi^{\frac{d}{2}} \Gamma(\Delta_i - \frac{d}{2})}  \frac{1}{ |\vec{x}-\vec{y}|^{2\Delta_i}}  \delta_{ij}  ~.
\end{equation}
For this specific computation, $I_{\rm ren}^{(0)}$ is the action in \eqref{eq:S_i}, plus counterterms, evaluated on-shell at tree-level. From \eqref{2pf} it is manifest that $\Delta_i$ is the conformal dimension of ${\cal O}_i$. To facilitate the notation in subsequent sections, we denote the normalization of the two-point function as
\begin{equation}\label{eq:2pf-norm}
    \mathsf{c}_{\Delta_i}\equiv  \frac{(2\Delta_i-d)\Gamma(\Delta_i)}{\pi^{\frac{d}{2}} \Gamma(\Delta_i - \frac{d}{2})}~.  
\end{equation}
Under these definitions, the connected $n$-point functions in the dual CFT is
\begin{equation}
    \langle {\cal O}_i(\vec{x}) \,\dots\, {\cal O}_j(\vec{y})\rangle \equiv-\frac{\delta I_{\rm ren}}{\delta \phi_i(\vec{x})\,\dots\,\delta \phi_j(\vec{y})}\;.
\end{equation}

As presented here, we are avoiding special values of the conformal dimensions. In particular, \eqref{2pf} applies without drama if
\begin{equation}\label{eq:Extra-Conditions-1}
    \Delta_i> \frac{d}{2}~.
\end{equation}
It is known how to modify the analysis to cover fields with $\Delta_i=\frac{d}{2}$, and the normalization \eqref{eq:2pf-norm} is modified in that case \cite{Freedman:1998tz}. For the sake of simplicity, we will avoid this special value in the general discussion. 
Our discussion will also include fields all the way down to the unitarity bound, i.e., between $\frac{d-2}{2}<\Delta_i^s<\frac{d}{2}$, and the majority of our analysis will apply in that range too. We will make the appropriate commentary when that range comes into play. Taking $\Delta_i^s=\frac{d-2}{2}$, at the unitarity bound, is also possible to analyze although delicate, and we will exclude it to avoid clutter.  

We will be specially interested in interactions among the bulk fields. In particular, we will consider the following three-point interaction between fields
\begin{equation}\label{int-d+1-generic}
 S_{\rm int}=-\lambda_{ijk} \int \dd^{d+1}x\,  \sqrt{g} \,\Phi_i(x)\,\Phi_j(x)\,\Phi_k(x)~,
\end{equation}
where $\lambda_{ijk}$ is a dimensionless coupling constant.\footnote{The action with the usual conventions of AdS/CFT is $- \lambda_{ijk} G_N^{1/2} \ell^{-2}\int \, \Phi_i\, \Phi_j\, \Phi_k$, where $G_N$ is Newton's constant and $\ell$ the AdS radius. With this convention the coupling constant $\lambda_{ijk}$ is dimensionless. } It is also possible to write interactions that include derivatives of the fields. We will be working to leading order in the coupling of the fields and focus only on the CFT three-point function, hence any number of derivatives that we add to \eqref{int-d+1-generic}  can be reabsorbed by a field redefinition of $\Phi_i$; see, for example, \cite{Gross:2017hcz}. Therefore \eqref{int-d+1-generic} is the most general cubic vertex to leading order in $\lambda_{ijk}$.   

A standard result in AdS/CFT is that the interaction term \eqref{int-d+1-generic} contributes to the three-point function of $\langle{\cal O}_i{\cal O}_j{\cal O}_k\rangle$ by the following integral 
\begin{equation}
    \begin{aligned}\label{Rastelli-Int-OG-1}
          A(\vec x_i,\vec x_j,\vec x_k)
    &=-\int \dd^{d+1}x\, \sqrt{g}\, K_{\Delta_i}(z_0,\vec z,\vec x_i) K_{\Delta_j}(z_0,\vec z,\vec x_j) K_{\Delta_k}(z_0,\vec z,\vec x_k)\\
    &=\frac{{\cal C}_{ijk}}{
    |x_i-x_j|^{\Delta_i+\Delta_j-\Delta_k}
    |x_j-x_k|^{\Delta_j+\Delta_k-\Delta_i}
    |x_k-x_i|^{\Delta_k+\Delta_i-\Delta_j}}~,
    \end{aligned}
\end{equation}
where $K_{\Delta}(z_0, \vec z,\vec x)$ is the bulk-to-boundary propagator \eqref{eq:bndy-bulk-prop} and 
\begin{equation}\label{Rastelli-Coef-OG-1}
    {\cal C}_{ijk}=-\frac{
    \Gamma\left(\frac{\Delta_i+\Delta_j-\Delta_k}{2}\right)
    \Gamma\left(\frac{\Delta_j+\Delta_k-\Delta_i}{2}\right)
    \Gamma\left(\frac{\Delta_k+\Delta_i-\Delta_j}{2}\right)}{2\pi^d \Gamma(\Delta_i-d/2) \Gamma(\Delta_j-d/2) \Gamma(\Delta_k-d/2)} \Gamma\left(\frac{\Delta_i+\Delta_j+\Delta_k-d}{2}\right)~.
\end{equation}
The convergence of the integrals involved relies on the triangle inequality, i.e., 
\begin{equation}
\Delta_i+\Delta_j >\Delta_k~, \qquad \forall ~  i\,,\,j\,,\,k~.    
\end{equation}
Provided the conformal dimensions involved do not correspond to a pole of ${\cal C}_{ijk}$, a finite answer can also be found via analytic continuation beyond the triangular inequality; see for example \cite{Fichet:2021pbn} for a discussion on these cases. In any case, our interest is precisely in the pathological values that sit at the poles of the Gamma functions.

%%%%%%%%%%%%%%%%%%%%%%%%%%%%%%%%%%%%%%%%%%%%%%%%%%%%%%%%%%%%%%%%%%%%%%%%%%%%%%%%%%%%%%%%
%%%%%%%%%%%%%%%%%%%%%%%%%%%%%%%%%%%%%%%%%%%%%%%%%%%%%%%%%%%%%%%%%%%%%%%%%%%%%%%%%%%%%%%%
%%%%%%%%%%%%%%%%%%%%%%%%%%%%%%%%%%%%%%%%%%%%%%%%%%%%%%%%%%%%%%%%%%%%%%%%%%%%%%%%%%%%%%%%
\section{Extremal cubic interactions in \texorpdfstring{AdS$_{d+1}$}{AdS}}\label{AdSd+1}
%%%%%%%%%%%%%%%%%%%%%%%%%%%%%%%%%%%%%%%%%%%%%%%%%%%%%%%%%%%%%%%%%%%%%%%%%%%%%%%%%%%%%%%%
%%%%%%%%%%%%%%%%%%%%%%%%%%%%%%%%%%%%%%%%%%%%%%%%%%%%%%%%%%%%%%%%%%%%%%%%%%%%%%%%%%%%%%%%
%%%%%%%%%%%%%%%%%%%%%%%%%%%%%%%%%%%%%%%%%%%%%%%%%%%%%%%%%%%%%%%%%%%%%%%%%%%%%%%%%%%%%%%%

Having established some of the basic setup and ingredients for massive scalar fields in AdS$_{d+1}$, the aim now is to quantify and interpret the effects of \eqref{int-d+1-generic} when $\Delta_k=\Delta_i+\Delta_j$, that is when we have an {\it extremal interaction}. Under these circumstances, it is evident from the divergence in \eqref{Rastelli-Coef-OG-1} that there is a problem with the integral in \eqref{Rastelli-Int-OG-1}. Here we will show how to extract the contribution to the three-point function  $\langle{\cal O}_i{\cal O}_j{\cal O}_k\rangle$  when an extremal interaction is non-trivial.   To keep the notation simple, we will consider three scalars $\Phi_i$, and consider an effective action on AdS$_{d+1}$ given by
\begin{equation}\label{OG-Action}
S=\sum_{i=1}^3 S_i +  S_{\rm ext}~, \qquad  S_{\rm ext}=-\lambda_{\rm ext} \int \dd^{d+1}x \sqrt{g} \,\Phi_1\,\Phi_2\,\Phi_3~.
\end{equation}
where $S_i$ is the free action \eqref{eq:S_i}, and the conformal dimension of the fields is such that
\begin{equation}\label{eq:extremal-delta}
    \Delta_3 = \Delta_1 +\Delta_2~,
\end{equation}
without loss of generality. We will call $\lambda_{\rm ext}$ in \eqref{OG-Action} an {\it extremal coupling}. 

In this section, we will focus on the contribution of \eqref{OG-Action} to the CFT three-point function to leading order in $\lambda_{\rm ext}$. To quantify these effects, we will use the asymptotic prescription along the lines of \cite{Freedman:1998tz,Skenderis:2002wp,vanRees:2011fr,vanRees:2011ir}. Here boundary conditions on the fields are set as an expansion near the boundary of AdS where the leading order is fixed, along the lines of \eqref{eq:leading-Phi}. Although this approach can have some drawbacks and ambiguities, we will show that it suffices here. To address these ambiguities, one can also consider a finite bulk cutoff where $z_0\in[\epsilon,\infty)$, with $0<\epsilon\ll1$, and construct appropriate counterterms to remove divergences as the cutoff reaches the boundary of AdS$_{d+1}$, see e.g. \cite{Mueck:1999nr,Botta-Cantcheff:2017qir}. We will illustrate how this works in App.\,\ref{sec:1+1=2} for a specific example, and find perfect agreement with the discussion here.

\subsection{Renormalization of the three-point function}\label{sec:three-point-extr}

For the interaction term \eqref{OG-Action}, the problematic contribution to the three-point function comes from the integral 
\begin{equation} \begin{aligned}\label{A-Int}
A_{\rm ext}(\vec{x}, \vec{y}, \vec{z}) &= - \int \frac{\dd^{d+1}w}{w_0^{d+1}}
K_{\Delta_1}(w_0,\vec w,\vec{x}) K_{\Delta_2}(w_0,\vec w,\vec{y}) K_{\Delta_3}(w_0,\vec w,\vec{z})~,
 \end{aligned}\end{equation}
where $ K_{\Delta_i}(w_0,\vec w,\vec{x})$ is the bulk-to-boundary propagator \eqref{eq:bndy-bulk-prop}.  The simplest way to proceed is to quantify the nature of the singularity in this integral. From here we will discuss how to extract the physically relevant information.

The first steps needed to manipulate \eqref{A-Int} follow from \cite{Freedman:1998tz}, which we will delineate in detail since certain aspects will be subtle. Following those traditional steps, we can first profit from the translation invariance on the boundary to put $\vec{z}=0$ and then perform an inversion transformation
\begin{equation}\label{Inversion-Trans}
    \vec{w}=\frac{\vec{w}'}{(w_0')^2+\vec{w}'^2}~, \qquad w_0=\frac{w_0'}{(w_0')^2+\vec{w}'^2}~,
\end{equation}
which is an isometry of \eqref{metric-d+1}. For the boundary points the inversion acts as
\begin{equation}\label{eq:Inversion-Trans-bndy}
    \vec x = \frac{\vec{x}'}{\vec{x}'^2}~.
\end{equation}
This greatly simplifies the integrand of \eqref{A-Int}, which now reads 
\begin{equation}
\begin{aligned}
    A_{\rm ext}(\vec{x},\vec{y},0) =&-
\frac{1}{|\vec{x}|^{2 \Delta_1}}
\frac{1}{|\vec{y}|^{2 \Delta_2}}
\prod_{i=1}^3\frac{\Gamma(\Delta_i)}{\pi^{\frac{d}{2}} \Gamma(\Delta_i-  \frac{d}{2})}\, R_0(\vec{x}',\vec{y}') ~,
\end{aligned}
\end{equation}
where 
\begin{equation}
    \begin{aligned}
     \label{eq:pre-R-Int}
         R_0(\vec{x}',\vec{y}')
&=\int_{0}^\infty \dd w'_0 \int \dd^d w' \frac{(w'_0)^a}{[(w'_0)^2+(\vec{w}'-\vec{x}')^2]^b[(w'_0)^2+(\vec{w}'-\vec{ y}')^2]^c}\;,
    \end{aligned}
\end{equation}
and we have introduced $a:=\Delta_1+\Delta_2+\Delta_3-(d+1)$, $b:=\Delta_1$, $c:=\Delta_2$. To consolidate the denominators, it is convenient to introduce a Feynman parameter in this integral. After performing the integral over $\vec{w}'$, we have
\begin{equation}
    \begin{aligned}
    \label{eq:R-Int-Analysis}
R_0(\vec{x}',\vec{y}') 
=\pi^{d/2}\frac{\Gamma(b+c-d/2)}{\Gamma(b)\Gamma(c)}\int_{0}^1 \dd u \int_{0}^\infty \dd w'_0  \frac{u^{b-1}(1-u)^{c-1}\;(w'_0)^a }{[(w'_0)^2+(1-u)u|\vec{x}'-\vec{y}'|^2]^{b+c-\frac d2}}~,
\end{aligned}
\end{equation}
where we have used \eqref{Feyn} and \eqref{RInt}. 

Up to this point, we have been very lenient and mainly assumed that $a$, $b$ and $c$ are positive and independent of each other; in particular, we have not used the extremality condition \eqref{eq:extremal-delta}.
However, \eqref{eq:R-Int-Analysis} will be divergent if we set $\Delta_3=\Delta_1 +\Delta_2$.\footnote{As mentioned in the introduction, there are other special values for which \eqref{eq:R-Int-Analysis} is ill-defined. Those cases will be discussed separately in Sec.\,\ref{sec:super-extremal} and Sec.\,\ref{sec:shadow-extremal}.} 
This is clear as we inspect the integral near the conformal boundary of AdS$_{d+1}$, i.e., for large values of $w_0'$: the integrand grows as $(w_0')^{\Delta_3-\Delta_2-\Delta_1-1}$, and this will give a logarithmic divergence at extremality. This UV divergence after inversion translates into an IR divergence in the original AdS$_{d+1}$ coordinates, and hence a UV divergence in the CFT$_d$.

To quantify the divergence at extremality, we will introduce an IR regulator in \eqref{eq:R-Int-Analysis}.\footnote{This is not the only regulator this computation admits. Still, it is important to emphasize that our conclusions do not depend on the choice of the regulator. For example, we could have taken $\Delta_3= \Delta_1+\Delta_2 +\epsilon$ and inspected the behavior of \eqref{eq:R-Int-Analysis} in the limit $\epsilon\to 0$. Another route is presented in App.\,\ref{sec:1+1=2}, where a finite cut-off is introduced in the bulk, see \cite{Mueck:1999nr,Botta-Cantcheff:2017qir}. All of these lead to the same conclusions regarding the non-trivial contribution to the three-point function. } The integral we will inspect is therefore 
\begin{equation}
R_\epsilon(\vec{x}',\vec{y}') := \pi^{d/2}\frac{\Gamma(b+c-d/2)}{\Gamma(b)\Gamma(c)}  \int_{0}^{1}\dd u\int_{0}^{1/\epsilon} \dd w'_0  \frac{u^{b-1}(1-u)^{c-1}\;(w'_0)^{a_{\rm ext}} }{((w'_0)^2+(1-u)u|\vec{x}'-\vec{y}'|^2)^{\frac{a_{\rm ext}+1}{2}}}~,
\end{equation}
where $0<\epsilon\ll1$ is the IR regulator, and we defined
\begin{equation}
    a_{\rm ext}:= 2\left(b+c\right) -d-1 ~,
\end{equation} 
which sets the extremality condition $\Delta_3=\Delta_1+\Delta_2$ in \eqref{eq:R-Int-Analysis}.
In order to avoid additional spurious divergences, we also take the restriction \eqref{eq:Extra-Conditions-1}. 
Carrying out the integral, we find
\begin{equation}
\begin{aligned}\label{eq:IR-reg-int}
\int_{0}^{1/\epsilon} \dd w'_0  \frac{(w'_0)^{a_{\rm ext}} }{[(w'_0)^2+(1-u)u|\vec{x}'-\vec{y}'|^2]^{\frac{a_{\rm ext}+1}{2} }}=&-\frac{1}{2} \log
   \left( |\vec{x'}-\vec{y'}|^2 u (1-u) \epsilon ^2 \right)\\ &\quad -\frac{1}{2}\psi\left(\frac{a_{\rm ext}+1}{2}\right) - \frac{\gamma}{2}    + O(\epsilon^2)~,
\end{aligned}    
\end{equation}
where $\psi(a)$ is the digamma function and $\gamma$ is Euler's constant. Further details of this integral are presented in App.\,\ref{app:conventions}, in particular, in \eqref{Aux-Ext1}-\eqref{Aux-Ext2}. Here we have expanded the result for small $\epsilon$ where in the last line we have dropped the convergent $O(\epsilon^2)$ terms. Any $O(\epsilon^0)$ contributions can be absorbed by rescaling $\epsilon$, therefore we get
\begin{equation}
    \begin{aligned}\label{eq:R-epsilon-final}
        R_\epsilon(\vec{x},\vec{y}) 
&=-\pi^{d/2}\frac{\Gamma(b+c-d/2)}{2 \Gamma (b+c)}\ln \left(|\vec{x'}-\vec{y'}|^2 \epsilon
   ^2\right)\\
   &=-\pi^{d/2}\frac{\Gamma(b+c-d/2)}{2 \Gamma (b+c)}\ln \left(\frac{|\vec{x}-\vec{y}|^2 }{|\vec{x}|^2|\vec{y}|^2}\epsilon
   ^2\right)~,
    \end{aligned}
\end{equation}
where in the last line we have reverted back to the original set of coordinates in \eqref{eq:Inversion-Trans-bndy}. Putting all the pieces together and recovering $\vec{z}$ with a rigid translation, we get
\begin{equation} \begin{aligned}
   A^\epsilon_{\rm ext}(\vec{x},\vec{y},\vec{z}) =\frac{1}{2\pi^d}
\frac{\mathsf{c}_{\Delta_1}\mathsf{c}_{\Delta_2}}{(2\Delta_1-d)(2\Delta_2-d)}
\frac{\ln \left(\frac{|\vec{x}-\vec{y}|^2 }{|\vec{x}-\vec{z}|^2|\vec{y}-\vec{z}|^2}\epsilon
   ^2\right)}{|\vec{x}-\vec{z}|^{2 \Delta_1}|\vec{y}-\vec{z}|^{2 \Delta_2}}~.
 \end{aligned}\end{equation}
This result needs a final adjustment, which is to renormalize away the $\epsilon$ dependence. This is made by improving the original action \eqref{OG-Action} with the anomalous counterterm 
\begin{equation}\label{General-e-CT}
    I_{\rm ct} = -\frac{\lambda_{\rm ext}}{2} \frac{\ln(\epsilon^2)}{(2\Delta_1-d)(2\Delta_2-d)} \int \dd^d z\,\sqrt{\gamma} \,\Pi_{\Phi_1}\Pi_{\Phi_2} \Phi_3 ~,
\end{equation}
where $\Pi_{\Phi_i}$ are the renormalized conjugate momentum of $\Phi_i$, i.e.,
\begin{equation}\label{def-Gen-Pi}
    \Pi_{\Phi_i}\equiv \frac{\delta S_{\rm ren}}{\delta \Phi_i}\;,
\end{equation}
and $S_{\rm ren}$ is the renormalized bulk action improved with all required counterterms to render the on-shell action finite. To leading order in the coupling, we have
\begin{equation}
 \lim_{z_0\to0}\Pi_{\Phi_i}(z_0,\vec x)=-z_0^{\Delta_i}\mathsf{c}_{\Delta_i} \int \dd^d z  \,\frac{\phi_i(\vec{z})}{ |\vec{z}-\vec{x}|^{2\Delta_i}} +O(\lambda_{\rm ext})~,
\end{equation}
where $\mathsf{c}_{\Delta_i}$ is given in \eqref{eq:2pf-norm}. Incorporating the counterterms, the renormalized vertex is then
\begin{equation}
    \begin{aligned}
   A^{\rm ren}_{\rm ext}(\vec{x},\vec{y},\vec{z}) &=\frac{1}{2}
\frac{\mathsf{c}_{\Delta_1}\mathsf{c}_{\Delta_2}}{(2\Delta_1-d)(2\Delta_2-d)}
\frac{\ln \left(\frac{|\vec{x}-\vec{y}|^2 }{|\vec{x}-\vec{z}|^2|\vec{y}-\vec{z}|^2}\right)}{|\vec{x}-\vec{z}|^{2 \Delta_1}|\vec{y}-\vec{z}|^{2 \Delta_2}}~. 
\end{aligned}
\end{equation}

With these ingredients, we can construct the renormalized on-shell action. Given our choice of counterterm in \eqref{General-e-CT}, we will have 
\begin{equation}
\begin{aligned}\label{eq:General-Ren-OnShell}
    I_{\rm ren}&= I_{\rm ren}^{(0)} + I_{\rm ren}^{(\rm int)} 
    +O(\lambda_{\rm ext}^2)~,
\end{aligned}    
\end{equation}
where 
\begin{equation}
\begin{aligned}\label{eq:General-Ren-OnShell-1}
    I_{\rm ren}^{(0)} &=-\frac 12\sum_{i=1}^3  \mathsf{c}_{\Delta_i} \int \, \phi_i({\vec x})\phi_i({\vec y})\left(  \frac{1}{ |\vec{x}-\vec{y}|^{2\Delta_i}}\right)~,\\
     I_{\rm ren}^{(\rm int)}  &= \frac{\lambda_{\rm ext}}{2}\frac{\mathsf{c}_{\Delta_1}\mathsf{c}_{\Delta_2}}{(2\Delta_1-d)(2\Delta_2-d)} \int \phi_1({\vec x})\phi_2({\vec y})\phi_3({\vec z}) \left(\frac{\ln \left(\frac{|\vec{x}-\vec{y}|^2 }{|\vec{x}-\vec{z}|^2|\vec{y}-\vec{z}|^2}\right)}{|\vec{x}-\vec{z}|^{2 \Delta_1}|\vec{y}-\vec{z}|^{2 \Delta_2}}\right)~.
\end{aligned}    
\end{equation}
Here $  I_{\rm ren}^{(0)}$ is the renormalized action of the free theory in \eqref{2pf} and $I_{\rm ren}^{(\rm int)}$ is three-point extremal interaction term to leading order in the coupling. We have omitted the differentials in the integrals above; it should be understood that the integrals are over the spatial coordinates $(\vec x,\vec y,\vec z)$ as appropriate. 
With this, the holographic three-point function to leading order is
\begin{equation}
    \begin{aligned}\label{General-3pf}
       \langle{\cal O}_1(\vec{x}) {\cal O}_2(\vec{y}) {\cal O}_3(\vec{z}) \rangle&\equiv-\frac{\delta I_{\rm ren}}{\delta \phi_1(\vec{x})\delta \phi_2(\vec{y})\delta \phi_3(\vec{z})}  \\
     &=-\frac{\lambda_{\rm ext}}{2} \frac{\mathsf{c}_{\Delta_1}\mathsf{c}_{\Delta_2}}{(2\Delta_1-d)(2\Delta_2-d)} \frac{\ln \left(\frac{|\vec{x}-\vec{y}|^2 }{|\vec{x}-\vec{z}|^2|\vec{y}-\vec{z}|^2}\right)}{|\vec{x}-\vec{z}|^{2 \Delta_1}|\vec{y}-\vec{z}|^{2 \Delta_2}}  ~.
    \end{aligned}
\end{equation}
It is important to stress that we are defining this three-point function when the three points do not coincide. The appearance of a logarithm and its interpretation in terms of a presumptive holographic CFT at the boundary will be more clear in the next subsection. 

It is essential to notice that from the bulk point of view, the putative counterterm
\begin{equation}\label{CT-constant-shift-1}
    \tilde I_{\rm ct}= c_{123}\; \frac{1}{\mathsf{c}_{\Delta_1}\mathsf{c}_{\Delta_2}}
   \int \dd^d z\sqrt{\gamma} \,\Pi_{\Phi_1}\Pi_{\Phi_2} \Phi_3 ~,
\end{equation}
with any finite coefficient $c_{123}$ is finite itself and can be freely added to the action \eqref{OG-Action} without altering the variational problem. These terms generate the same effect in the correlators as the boundary terms used in \cite{DHoker:1999jke,Arutyunov:2000ima} and introduce scheme-dependent contributions, i.e., terms that can be modified by $\epsilon$ rescalings \cite{Bzowski:2015pba}. Terms of the form \eqref{CT-constant-shift-1} shift \eqref{General-3pf} by 
\begin{equation}\label{CT-constant-shift-2}
    \langle{\cal O}_1(\vec{x}) {\cal O}_2(\vec{y}) {\cal O}_3(\vec{z}) \rangle \to  \langle{\cal O}_1(\vec{x}) {\cal O}_2(\vec{y}) {\cal O}_3(\vec{z}) \rangle +
    \frac{c_{123}}{|\vec{x}-\vec{z}|^{2 \Delta_1}|\vec{y}-\vec{z}|^{2 \Delta_2}}~,
\end{equation}
which are contributions compatible with conformal symmetry.
Our result in \eqref{General-3pf} is the contribution to the correlator that cannot be removed in any renormalization scheme. 

\subsection{Anomalous dimensions}

The presence of the anomalous counterterm \eqref{General-e-CT} and the logarithmic contributions to CFT$_d$ correlation functions \eqref{General-3pf} deserve further inspection.  One clear effect of \eqref{General-e-CT} is to introduce an anomaly in the boundary stress tensor, and a rushed conclusion would be to declare that this theory is breaking conformal invariance. In this portion, we will address the interpretation of this counterterm by inspecting two-point functions, which will account correctly for the logarithmic contributions due to operator mixing in the CFT.

We start by taking ${\cal O}_1(\vec x)$ close to ${\cal O}_2 (\vec y)$ in correlation functions, that is, by looking at the behavior of the composite operator
\begin{equation}
 {\cal O}_{1+2}(\vec{x})\equiv\;:\!{\cal O}_1 {\cal O}_2\!:\!(\vec{x}) ~.
\end{equation}
Outside of extremality, it is well-known how to analyze and interpret the limit $|\vec{x}-\vec{y}|\to0$ from the three-point function in terms of an OPE expansion.  
For general $\Delta_i$, one can readily check that regularization of the CFT three-point function \eqref{Rastelli-Int-OG-1} leads to a vanishing contact contribution whenever any two of the three points meet. In our case, $I_{\rm ren}^{(\rm int)}$ in \eqref{eq:General-Ren-OnShell-1} receives a non-vanishing contribution as $\vec x\to \vec y$. To obtain such a contact term we can take the limit $\vec y'\to \vec x'$ in \eqref{eq:pre-R-Int}, in which the denominator becomes a single factor. From there, a similar bulk IR regulation can be introduced producing a finite distance contribution. The counterterm \eqref{General-e-CT} alone is enough to render finite such a vertex. We find 
\begin{equation}
\begin{aligned}\label{eq:Iren-c}
   I_{\rm ren}^{(\rm c)} = -\lambda_{\rm ext} \frac{\mathsf{c}_{\Delta_1}\mathsf{c}_{\Delta_2}}{(2\Delta_1-d)(2\Delta_2-d)} \int \, \phi_{1+2}(\vec{x})\phi_3({\vec z})\left( \frac{\ln \left(|\vec{x}-\vec{z}|^2\right)}{|\vec{x}-\vec{z}|^{2 \Delta_1 +2\Delta_2}}\right)~, 
\end{aligned}    
\end{equation}
where we defined the $\phi_{1+2}$ notation to indicate a source for the $:\!{\cal O}_1 {\cal O}_2\!:$ operator.
This implies
\begin{equation}
\begin{aligned}\label{General-Mixed-2pf}
    \langle {\cal O}_{1+2}(\vec{x}){\cal O}_3(\vec{z}) \rangle&\equiv-\frac{\delta I_{\rm int}^{\rm ren}}{\delta \phi_{1+2}(\vec{x})\delta \phi_3(\vec{z})}\\
    &=\lambda_{\rm ext}\frac{\mathsf{c}_{\Delta_1}\mathsf{c}_{\Delta_2}}{(2\Delta_1-d)(2\Delta_2-d)}  \frac{\ln \left(|\vec{x}-\vec{z}|^2\right)}{|\vec{x}-\vec{z}|^{2 \Delta_1 +2 \Delta_2}}~.
\end{aligned}
\end{equation}
Not surprisingly we find a mixing between operators. This is a consequence of having a degenerate spectrum at $\lambda_{\rm ext}=0$, where ${\cal O}_3$ and ${\cal O}_{1+2}$ have the same conformal dimension. Adding an interaction term, $\lambda_{\rm ext}\neq0$, lifts this degeneracy which is reflected by the non-zero answer in \eqref{General-Mixed-2pf}.  

To see this explicitly let us introduce a new basis of operators
\begin{equation} \begin{aligned}\label{O+-Def}
   {\cal O}_{\pm}(\vec x)&\equiv \frac{1}{\sqrt{2}}\left({\cal O}_{3}(\vec x) \pm \tilde c \, {\cal O}_{1+2}(\vec x)\right)~,
 \end{aligned}\end{equation}
with $\tilde c \in \mathbb{R}$. Making this basis orthogonal gives 
\begin{equation}\label{No-Mix+-}
    \langle {\cal O}_{+}(\vec x){\cal O}_{-}(0)\rangle =0 \qquad \Rightarrow \qquad \tilde c = \sqrt{\frac{\langle {\cal O}_{3}(\vec x){\cal O}_{3}(0)\rangle}{\langle {\cal O}_{1+2}(\vec x){\cal O}_{1+2}(0)\rangle}}~.
\end{equation}
To leading order in the coupling, we normalize the two-point function of ${\cal O}_3$ as \eqref{2pf}. For ${\cal O}_{1+2}$ we will fix the normalization from the 4-point function by making the points coincide:
\begin{equation}
\begin{aligned}\label{Ext-point-norm}
    \lim_{\subalign{\vec w\to \vec y \\ \vec z\to \vec x }}\;
    \langle {\cal O}_{1}(\vec x){\cal O}_{2}(\vec z){\cal O}_{1}(\vec y){\cal O}_{2}(\vec w)\rangle &\sim 
    \langle {\cal O}_{1}(\vec x){\cal O}_{1}(\vec y)\rangle \langle{\cal O}_{2}(\vec x){\cal O}_{2}(\vec y)\rangle
    \sim 
      \frac{\mathsf{c}_{\Delta_1}\mathsf{c}_{\Delta_2}}{ |\vec{x}-\vec{y}|^{2\Delta_1+2\Delta_2}}~.
\end{aligned}    
\end{equation}
The leading terms are  Wick contractions of generalized free fields, and this will set the normalization of correlators involving composite operators.\footnote{This correlation function can be found from $e^{-I_{\rm ren}}$ since the leading contribution is disconnected.}
With this we have 
\begin{equation}\label{eq:defn-ctilde}
    \tilde c =
    \sqrt{\frac{\mathsf{c}_{\Delta_1+\Delta_2}}{\mathsf{c}_{\Delta_1}\mathsf{c}_{\Delta_2}}}~.
\end{equation}
Next, if we now inspect the two-point function of ${\cal O}_\pm$ we have
\begin{equation}\label{2pf+-}
    \begin{aligned}
        \langle {\cal O}_{\pm}(\vec x){\cal O}_{\pm}(\vec z)\rangle & =  \langle {\cal O}_{3}(\vec x){\cal O}_{3}(\vec z)\rangle \Big(1 \pm \tilde\gamma\ln \left(|\vec{x}-\vec{z}|^2\right)+O(\lambda_{\rm ext}^2)\Big)\\
    & =   \frac{\mathsf{c}_{\Delta_1+\Delta_2}}{ |\vec{x}-\vec{z}|^{2(\Delta_1+\Delta_2\mp \tilde\gamma)}}+O(\lambda_{\rm ext}^2)~,
    \end{aligned}
\end{equation}
where
\begin{equation} \begin{aligned}\label{Gen-Anom}
\tilde\gamma&=\frac{\lambda_{\rm ext}}{\pi^d}\frac{ \Gamma \left(\Delta _1\right) \Gamma \left(\Delta _2\right)}{\Gamma \left(\Delta
   _1-\frac{d}{2}\right) \Gamma \left(\Delta _2-\frac{d}{2}\right)} \frac{1}{\sqrt{\mathsf{c}_{\Delta_1+\Delta_2}\mathsf{c}_{\Delta_1}\mathsf{c}_{\Delta_2} }}\\&=\frac{\lambda_{\rm ext} }{(2\Delta_1-d)(2\Delta_2-d)}\frac{1}{\tilde c}~,
 \end{aligned}\end{equation}
is an anomalous dimension that corrects the conformal dimension of the operators ${\cal O}_\pm$ to 
\begin{equation}\label{Delta-pm}
    \Delta_\pm = \Delta_1+\Delta_2 \mp \tilde\gamma~,
\end{equation}
in the presence of the extremal coupling $\lambda_{\rm ext}$. 

We can also reproduce $\tilde\gamma$ by looking at the three-point function. Using \eqref{General-3pf}, \eqref{O+-Def} and \eqref{Ext-point-norm}, we obtain  
\begin{equation}
    \langle {\cal O}_1({\vec x}){\cal O}_2({\vec y}){\cal O}_\pm({\vec z})\rangle = \pm\sqrt{\frac{\mathsf{c}_{\Delta_1+\Delta_2}\, \mathsf{c}_{\Delta_1}\mathsf{c}_{\Delta_2}}{2} }
     \frac{1}{|\vec{x}-\vec{y}|^{\pm\tilde \gamma}|\vec{x}-\vec{z}|^{2 \Delta_1 \mp\tilde\gamma}|\vec{y}-\vec{z}|^{2 \Delta_2\mp\tilde\gamma}}  ~,
\end{equation}
with $\tilde \gamma$ as in \eqref{Gen-Anom}, and hence in agreement with \eqref{2pf+-}.

This analysis shows that the logarithmic contribution to the correlation function is tied to lifting a degenerate spectrum when we turn on an extremal coupling. To this order in perturbation theory, ${\cal O}_{\pm}$ are the appropriate basis of operators which should be used in the presence of this coupling.

%%%%%%%%%%%%%%%%%%%%%%%%%%%%%%%%%%%%%%%%%%%%%%%%%%%%%%%%%%%%%%%%%%%%%%%%%%%%%%%%%%%%%%%%
%%%%%%%%%%%%%%%%%%%%%%%%%%%%%%%%%%%%%%%%%%%%%%%%%%%%%%%%%%%%%%%%%%%%%%%%%%%%%%%%%%%%%%%%
%%%%%%%%%%%%%%%%%%%%%%%%%%%%%%%%%%%%%%%%%%%%%%%%%%%%%%%%%%%%%%%%%%%%%%%%%%%%%%%%%%%%%%%%
\section{Other peculiar cubic interactions in \texorpdfstring{AdS$_{d+1}$}{AdS}}\label{sec:peculiar}
%%%%%%%%%%%%%%%%%%%%%%%%%%%%%%%%%%%%%%%%%%%%%%%%%%%%%%%%%%%%%%%%%%%%%%%%%%%%%%%%%%%%%%%%
%%%%%%%%%%%%%%%%%%%%%%%%%%%%%%%%%%%%%%%%%%%%%%%%%%%%%%%%%%%%%%%%%%%%%%%%%%%%%%%%%%%%%%%%
%%%%%%%%%%%%%%%%%%%%%%%%%%%%%%%%%%%%%%%%%%%%%%%%%%%%%%%%%%%%%%%%%%%%%%%%%%%%%%%%%%%%%%%%

In this section, we explore other ``peculiar'' interactions between massive scalar fields in AdS$_{d+1}$ for which the naive tree-level three-point function \eqref{Rastelli-Coef-OG-1} contains divergences. There are two cases we will consider, which correspond to cubic interactions of the form \eqref{int-d+1-generic}, where the conformal dimensions of the fields obey  
\begin{equation}
    \begin{aligned}\nn
\textrm{{\it Super-extremal}:}& \qquad       \Delta_i+\Delta_j=\Delta_k-2n~, \quad n \in \mathbb{N}~, \\
\textrm{{\it Shadow-extremal}:}& \qquad       \Delta_i+\Delta_j+\Delta_k=d~.
    \end{aligned}
\end{equation} 
In the following, we will show how to obtain a renormalized on-shell action for each case and discuss the resulting three-point functions. 
For both scenarios, we will see  similarities with the extremal coupling in Sec.\,\ref{AdSd+1}, and we will make the appropriate comparisons.  

We should mention that the special cases
\begin{equation}\nn
    \Delta_i +\Delta_j + \Delta_k= d- 2 n ~, \qquad n\in \mathbb{N}~,
\end{equation}
also lead to divergences. The underlying mathematics needed to regulate this divergence is very similar to the logic used for extremal interactions. However, unitarity heavily restricts the number of physically relevant examples that comply with this condition. Shadow-extremal couplings ($n=0$) themselves can only happen in $d\leq6$ and super-shadow-extremal couplings are forbidden for $n>1$ and only allowed for $n=1$ for $d\leq2$. As such, we do not discuss these last cases further.

\subsection{Super-extremal couplings}\label{sec:super-extremal}

In this peculiar case, we will consider again three massive scalar fields $\Phi_i$, and an effective action on AdS$_{d+1}$ given by
\begin{equation}\label{OG-Action-se}
S=\sum_{i=1}^3 S_i +  S_{\rm se}~, \qquad  S_{\rm se}=-\lambda_{\rm se} \int \dd^{d+1}x \sqrt{g} \,\Phi_1\,\Phi_2\,\Phi_3~.
\end{equation}
where $S_i$ is the free action \eqref{eq:S_i}, and the conformal dimension of the fields is such that
\begin{equation}\label{eq:super-delta}
    \Delta_3 = \Delta_1 +\Delta_2 +2n~, \quad n\in \mathbb{N}~,
\end{equation}
without loss of generality. The coupling constant $\lambda_{\rm se}$ in \eqref{OG-Action-se} will be referred to as a {\it super-extremal coupling}. One can see that our analysis up to \eqref{eq:R-Int-Analysis} follows identically as for the extremal case. Regulating the integral once again with a cutoff in $w_0'$, i.e.,  $w_0'\in[0,1/\epsilon]$ with $0<\epsilon\ll1$,  leads to power-law as well as logarithmic divergences in the cut-off that can all be removed via counterterms;  this leads to a renormalized on-shell action. The procedure of Sec.\,\ref{sec:three-point-extr} applied here then tell us that the appropriate counterterm is given by 
\begin{equation}
    \begin{aligned}\label{SE-ct}
    I_{\rm ct} &=\lambda_{\rm se}\,\ln(\epsilon^2)\,\frac{ (-1)^n\Gamma\left(\Delta _1+n\right) \Gamma
   \left(\Delta _2+n\right) \Gamma\left(\Delta _1+\Delta_2+n-\frac d2\right)}{2  \pi ^{d} n!\Gamma \left(\Delta_1-\frac{d}{2}\right) \Gamma\left(\Delta _2-\frac{d}{2}\right)\Gamma\left(\Delta_1+\Delta _2+2n-\frac d2\right)}  \\
   &\qquad\qquad\times \int \frac{|\vec x-\vec y|^{2 n}}{ |\vec x-\vec z|^{2(\Delta_1+n)} |\vec y-\vec z|^{2(\Delta_2+n)} } \phi_1(\vec x)\phi_2(\vec y)\phi_3(\vec z)\\
   &= \lambda_{\rm se}\,\ln(\epsilon^2)\,\frac{(-1)^n \Gamma \left(n+\Delta_1\right) \Gamma \left(n+\Delta_2\right) \Gamma
   \left(\Delta _1+\Delta_2+n-\frac{d}{2}\right)}{2 n! \left(d-2 \Delta_1\right) \left(d-2 \Delta _2\right)
   \Gamma\left(\Delta _1\right) \Gamma\left(\Delta _2\right) \Gamma
   \left(\Delta _1+\Delta_2+2n-\frac{d}{2}\right)}  \\
   &\qquad\qquad\times\int \dd^d z \sqrt{\gamma} \,\Phi_3(\vec z) \left(\prod_{i=0}^{n} \square^{\Delta_1+i}_{\Delta_2+i}[\Pi_{\Phi_1}(\vec z),\Pi_{\Phi_2}(\vec z)]\right)\,,
    \end{aligned}
\end{equation}
where in the second line we have rewritten the counterterm in covariant form with the aid of the ad-hoc notation for the bilinear operator
\begin{equation}
\begin{aligned}
    \label{bil-square-def}
    \square^{\Delta_1+i}_{\Delta_2+i}[a_1(\vec z),a_2(\vec z)]\equiv \frac{a_2(\vec z)\square_\gamma a_1(\vec z)+a_1(\vec z)\square_\gamma a_2(\vec z)}{(2(\Delta_1+i))^2}&-2\gamma^{\mu\nu}\,\frac{\partial_\mu a_1(\vec z)}{2(\Delta_1+i)}\frac{\partial_\nu a_2(\vec z)}{2(\Delta_2+i)}~,
\end{aligned}
\end{equation}
where recall that $\gamma^{ab}=\epsilon^2 \delta^{ab}$ is the metric at the boundary and $\square_\gamma=\epsilon^2\partial_a\partial_a$ is the  Laplacian induced at the boundary. The operator above is defined so that when acting on the conjugate momenta $\Pi_{\Phi_i}(\vec x)$ it yields
\begin{equation}
\prod_{i=0}^{n} \square^{\Delta_1+i}_{\Delta_2+i}[\Pi_{\Phi_1}(\vec z),\Pi_{\Phi_2}(\vec z)]= \epsilon^{2n}\frac{|\vec x-\vec y|^{2 n}}{ |\vec x-\vec z|^{2n} |\vec y-\vec z|^{2n} }\Pi_{\Phi_1}(\vec z)\,\Pi_{\Phi_2}(\vec z)~.
\end{equation}
One can rewrite the counterterm \eqref{SE-3pf} in many ways by using integration by parts, but we have chosen its form to compare it with the extremal case in \eqref{General-e-CT}.

The renormalized action $I_{\rm se}^{\rm ren}$ can be constructed by the steps in Sec.\,\ref{sec:three-point-extr} and using the counterterm \eqref{SE-ct}. The three-point function at finite distance, to leading order in $\lambda_{\rm se}$, reads
\begin{equation}
    \begin{aligned}\label{SE-3pf}
       \langle{\cal O}_1(\vec{x}) {\cal O}_2(\vec{y}) {\cal O}_3(\vec{z}) \rangle_{\rm se}&\equiv-\frac{\delta I_{\rm se}^{\rm ren}}{\delta \phi_1(\vec{x})\delta \phi_2(\vec{y})\delta \phi_3(\vec{z})}  \\
    &=\lambda_{\rm se}\frac{ (-1)^n \Gamma\left(\Delta_1+n\right) \Gamma\left(\Delta_2+n\right) \Gamma\left(\Delta_1+\Delta_2+n-\frac d2\right)}{2\pi^{d} n!\Gamma \left(\Delta_1-\frac{d}{2}\right) \Gamma\left(\Delta _2-\frac{d}{2}\right)\Gamma \left(\Delta_1+\Delta _2+2n-\frac d2\right)}
   \\
   &\qquad\qquad\times\frac{|\vec x-\vec y|^{2 n}\ln \left(\frac{|\vec x-\vec y|^2}{|\vec x-\vec z|^2
   |\vec y-\vec z|^2}\right)}{ |\vec x-\vec z|^{2(\Delta_1+n)} |\vec y-\vec z|^{2(\Delta_2+n)} }~.
    \end{aligned}
\end{equation}

Much like the extremal case, the logarithmic divergences reveal an anomalous dimension in the spectrum. However, for a cubic super-extremal bulk coupling of order $n$, there are now $n(n+1)/2$ CFT operators of the same dimension of ${\cal O}_3$, which are all possible scalar descendants of ${\cal O}_{1+2}$. 
However, only one of these gets mixed up with ${\cal O}_3$, at least to leading order in $\lambda_{\rm se}$. This can be seen from the fact that there is a single finite distance logarithmic correlator resulting from $\lambda_{\rm se}\neq0$. The operator coupled to ${\cal O}_3$ can be spotted by profiting from translation invariance by coming back to \eqref{SE-3pf}, to obtain
\begin{equation}
    \begin{aligned}
\langle :\!\!\partial^{2n}{\cal O}_1 {\cal O}_2\!\!:\!\!(0) {\cal O}_3(\vec{x}) \rangle_{\rm se} & \equiv \lim_{\vec y\to0}\langle(\nabla_{y}^2)^n{\cal O}_1(\vec{y}) {\cal O}_2(0) {\cal O}_3(\vec{x}) \rangle_{\rm se}  \\
&=\lambda_{\rm se}\frac{ (-1)^n \Gamma
\left(\Delta _1+n\right) \Gamma
\left(\Delta _2+n\right) \Gamma
\left(\Delta _1+\Delta_2+n-\frac d2\right)}{2\pi^{d} n!\Gamma \left(\Delta_1-\frac{d}{2}\right) \Gamma
\left(\Delta _2-\frac{d}{2}\right)
\Gamma \left(\Delta_1+\Delta _2+2n-\frac d2\right)} \\
&\qquad\qquad\times\frac{-2(2n)!\ln \left(|\vec x|^{2}\right)}{ |\vec x|^{2(\Delta_1+\Delta_2+2n)}}~.
\end{aligned}
\end{equation}
The anomalous dimension computation now follows an analogous path as in the extremal case and the general result is not very illuminating, involving combinatoric coefficients related to the normalization of the descendants of the $\Delta_1$ and $\Delta_2$ primaries. Thus, we leave this computation implicit.

\subsection{Shadow-extremal couplings}\label{sec:shadow-extremal}

Finally, consider again theory with an extremal coupling as in \eqref{OG-Action}, where now one or more bulk fields lie in the double quantization window \cite{Witten:2001ua}. In this scenario, one might try to avoid the logarithmic correlators from Sec.\,\ref{sec:three-point-extr} by quantizing in the alternative quantization scheme, i.e., the dual operator for the appropriate field will now have $\Delta^{s}_i=d-\Delta_i$. We now study this set-up, where now the extremal interaction will turn into a shadow-extremal one.

Before doing so, it is important to discuss some of the properties of the shadow-extremal degeneracy and its tension with unitarity. In particular, notice that the condition
\begin{equation}\label{she-Action-1}
     \Delta_1 +\Delta_2 +\Delta_3=d~.
\end{equation}
is forbidden by our definition of $\Delta_i> d/2$ in \eqref{eq:Extra-Conditions-1}, but it can be met if one or more of the $\Delta_i$ are replaced by $\Delta_i^s<d/2$. However, a physical unitary operator with $\Delta_i^s<d/2$ quickly becomes in tension with the unitarity bound $(d-2)/2<\Delta_i^s$ in the CFT$_d$. One can readily show, for example, that the shadow-extremal degeneracy for a cubic bulk vertex only consistent with CFT$_d$ unitarity bounds in $d\leq6$.
Without loss of generality, let us arrange the masses of the fields such that
\begin{equation}\label{she-Action-2}
d-\Delta_3 = \Delta_3^s \leq \Delta_2 \leq \Delta_1 \;, \qquad \qquad \Delta_3 = \Delta_1 +\Delta_2~.
\end{equation}
Since $\Delta_{1,2}>d/2$, the equations above force $\Delta_3^s<0$, hence the QFT on AdS$_{d+1}$ is unstable and incompatible with unitarity in the CFT. This implies that any unitary example of a shadow-extremal cubic bulk interaction requires at least two of the three fields to have masses in the double quantization window.  

However, we will ignore constraints from unitarity bounds to show how a bulk shadow-extremal coupling leads to operator mixing and anomalous dimensions. We will therefore study a toy model where 
\begin{equation}\label{she-Action-3}
 \Delta_1 +\Delta_2 +\Delta_3^s= d \qquad\qquad \Leftrightarrow \qquad\qquad \Delta_3=\Delta_1+\Delta_2~.
\end{equation}
We stress that this is a QFT on AdS$_{d+1}$ that is in tension with unitarity, and therefore with limited scope if pushed further. An example compatible with unitarity is presented in App.\,\ref{app:1/3+1/3+1/3=1} showing that our conclusions here are physically sound regarding the anomalies we see in the on-shell action.

Our choice of spectrum \eqref{she-Action-2}, and its cubic interaction, is described by the same bulk action as for the extremal case defined in \eqref{OG-Action}, which is
\begin{equation}\label{she-Action}
S=\sum_{i=1}^3 S_i +  S_{\rm she}~, \qquad  S_{\rm she}=-\lambda_{\rm she} \int \dd^{d+1}x \sqrt{g} \,\Phi_1\,\Phi_2\,\Phi_3~,
\end{equation}
where $S_i$ is the free action \eqref{eq:S_i}. The difference here is that we will now quantize the field $\Phi_3$ in the alternative quantization scheme. For that reason, we have relabeled $\lambda_{\rm ext} \to \lambda_{\rm she}$ to differentiate between the extremal and shadow-extremal cases. 
The coupling constant $\lambda_{\rm she}$ in \eqref{she-Action} will be referred to as {\it shadow-extremal coupling}. 
We will keep $\phi_3(\vec x)$ as the notation for the source corresponding to the standard quantization scheme, but now interpret the bulk field $\Phi_3$ as having a dual operator ${\cal O}^s_3$ of conformal dimension $\Delta^s_3=d-\Delta_3$ with source $\phi_3^s(\vec x)$.

Having the same bulk action as in Sec.\,\ref{AdSd+1}, the regularization of the shadow-extremal case then follows an analogous path as the extremal case, up to the renormalized on-shell action 
\eqref{eq:General-Ren-OnShell}. 
At that point, one needs to functionally invert the relation between the sources using standard techniques of the AdS/CFT dictionary \cite{Witten:2001ua,Freedman:2016yue}. 
The alternative quantization scheme is defined via a change of boundary data in \eqref{eq:General-Ren-OnShell} by complementing the on-shell action $I_{\rm ren}$ with
\begin{equation}\begin{aligned}\label{I-she-ren-0}
     I_{\rm ren} \to I_{\rm she}^{\rm ren}&\equiv I_{\rm ren}+\int \dd^dx\sqrt{\gamma}\, \Pi_{3}\,\Phi_3\\ &= I_{\rm ren}+\int\dd^dx \,\phi^s_{3}(\vec x)\,\phi_{3}(\vec x) ~. 
\end{aligned}
\end{equation}
Since $I_{\rm ren}$ itself defined a well posed variational problem in terms of $\phi_{3}$ (keeping $\phi_1$ and $\phi_2$ fixed), it follows that
\begin{equation}
    \delta\left(I_{\rm ren}+\int\phi^s_{3}\,\phi_{3} \right)=\int \left(\frac{\delta I_{\rm ren}}{\delta \phi_{3}}+\phi^s_{3}\right)\delta \phi_{3}+\int \phi_{3}\delta\phi^s_{3}~.
\end{equation}
Thus, a well-posed variational problem is defined in terms of $\phi^s_{3}$ as long as 
\begin{equation}\label{phi-minus-def}
    \phi^s_{3}(\vec x)\equiv - \frac{\delta I_{\rm ren}[\phi_{3}]}{\delta\phi_{3}(\vec x)} ~,
\end{equation}
where this equation is to be functionally inverted to obtain $\phi_3[\phi_3^s(\vec x)]$. The result is non-local, and we will be solving it order by order in $\lambda_{\rm she}$ and the number of sources, which we collectively denote $O(\phi^n)$ with integer $n$. To leading order we therefore find 
\begin{equation} \begin{aligned}
    \phi_3[\phi_3^s(\vec z)]=&\frac{\mathsf{c}_{\Delta^s_3}}{(2\Delta^s_3-d)^2 }\int \dd^d w \, \phi_3^s(\vec w)
    \left(\frac{1}{|\vec z - \vec w|^{2\Delta^s_3}}\right) 
     \\ 
    &+ \frac{\lambda_{\rm she}}{4\pi^d} \frac{ \Gamma
   \left(\frac{d}{2}-\Delta _1\right)
   \Gamma \left(\frac{d}{2}-\Delta
   _2\right) \Gamma
   \left(\Delta _1+\Delta
   _2-\frac{d}{2}\right)}{\Gamma \left(\Delta
   _1-\frac{d}{2}\right) \Gamma \left(\Delta
   _2-\frac{d}{2}\right)\Gamma
   \left(\frac{d}{2}-\Delta _1-\Delta
   _2+1\right) }  \\
   &\qquad\qquad\times \int  \,\phi_1(\vec x)\phi_2(\vec y) \,\left( \frac{\ln \left(\frac{|\vec x-\vec y|^2 }{|\vec x-\vec z|^2 |\vec y-\vec z|^2}\right)}{|\vec x-\vec z|^{d-2\Delta_2}|\vec y-\vec z|^{d-2\Delta _1}|\vec x-\vec y|^{2(\Delta _1+\Delta _2)-d }}\right) \\
   &+ O(\lambda_{\rm she}^2)+ O(\phi^3)~.
 \end{aligned}\end{equation}
The on-shell action defined in \eqref{I-she-ren-0} then becomes
\begin{equation}
\begin{aligned}\label{SHE-Ren-OnShell}
    I_{\rm she}^{\rm ren}=&-\sum_{i=1}^2 \frac{\mathsf{c}_{\Delta_i}}{2} \int \, \phi_i({\vec x}) \phi_i({\vec y}) \left(\frac{1}{ |\vec{x}-\vec{y}|^{2\Delta_i}}\right)\\&- \frac{\mathsf{c}_{\Delta_3^s}}{2(2\Delta_3^s-d)^2}\int \phi_3^s({\vec x})\phi_3^s({\vec y})\left(\frac{1}{|\vec x - \vec y|^{2\Delta_3^s}}\right)\\
    &+ \frac{\lambda_{\rm she} }{2\pi^d}\frac{ \Gamma\left(\frac{d}{2}-\Delta _1\right)   \Gamma \left(\frac{d}{2}-\Delta_2\right) \Gamma
   \left(\Delta _1+\Delta_2-\frac{d}{2}\right)}{\Gamma \left(\Delta_1-\frac{d}{2}\right) \Gamma \left(\Delta_2-\frac{d}{2}\right)\Gamma\left(\frac{d}{2}-\Delta _1-\Delta_2+1\right) }  \\
   &\qquad\qquad\times  \int  \phi_1({\vec x})\phi_2({\vec y})\phi^s_3({\vec z}) \left(\frac{\ln \left(\frac{|\vec x-\vec y|^2 }{|\vec x-\vec z|^2 |\vec y-\vec z|^2}\right)}{|\vec x-\vec z|^{d-2\Delta_2}|\vec y-\vec z|^{d-2\Delta _1}|\vec x-\vec y|^{2(\Delta _1+\Delta _2)-d }}\right) \\
   &+ O(\lambda_{\rm she}^2)+ O(\phi^4)\;.
\end{aligned}    
\end{equation}

The two-point function for ${\cal O}_3^s$, at leading order, reads 
\begin{equation} \begin{aligned}
     \langle {\cal O}_3^s(\vec{x}) {\cal O}_3^s(\vec{y}) \rangle=\frac{1}{(2\Delta_3^s-d)^2}\times\left(\frac{\mathsf{c}_{\Delta_3^s}}{|\vec x - \vec y|^{2\Delta_3^s}}\right)~.
 \end{aligned}\end{equation} 
This differs from the normalization used for ${\cal O}_3$ due to inverting the boundary conditions, which is standard in AdS \cite{Mueck:1999nr,Freedman:2016yue}. 
For the three-point function, we get
\begin{equation}
    \begin{aligned}\label{SHE-3pf}
       \langle{\cal O}_1(\vec{x}) {\cal O}_2(\vec{y}) {\cal O}_3^s(\vec{z}) \rangle&\equiv-\frac{\delta I_{\rm she}^{\rm ren}}{\delta \phi_1(\vec{x})\delta \phi_2(\vec{y})\delta \phi_3^s(\vec{z})}  \\
    &=-\frac{\lambda_{\rm she}}{2 \pi ^{d}}\,\frac{ \Gamma
   \left(\frac{d}{2}-\Delta _1\right)
   \Gamma \left(\frac{d}{2}-\Delta
   _2\right) \Gamma
   \left(\Delta _1+\Delta
   _2-\frac{d}{2}\right)}{\Gamma \left(\Delta
   _1-\frac{d}{2}\right) \Gamma \left(\Delta
   _2-\frac{d}{2}\right)\Gamma
   \left(\frac{d}{2}-\Delta _1-\Delta
   _2+1\right) }
   \\
   &\qquad\qquad\times\frac{
   \ln \left(\frac{|\vec x-\vec y|^2 }{|\vec x-\vec z|^2 |\vec y-\vec z|^2}\right)}{|\vec x-\vec z|^{d-2\Delta
   _2}|\vec y-\vec z|^{d-2\Delta_1}|\vec x-\vec y|^{2(\Delta_1+\Delta _2)-d }} ~,
    \end{aligned}
\end{equation}
The mixed two-point function can be obtained, as we did for eq. \eqref{eq:Iren-c}, by regularization of the three-point function whenever the points meet.\footnote{See App.\,\ref{app:1/3+1/3+1/3=1} for more details of this limiting procedure.} We get,
\begin{equation}
    \begin{aligned}\label{SHE-2pf-mix}
       \langle{\cal O}_{1+2}(\vec{x}) {\cal O}_3^s(\vec{y}) \rangle&\equiv-\frac{\delta I_{\rm she}^{\rm ren}}{\delta \phi_{1+2}(\vec{x})\delta \phi_3^s(\vec{y})} = \lambda_{\rm she}  \frac{\mathsf{c}_{1+2}^s }{|\vec x-\vec y|^d}~,
    \end{aligned}
\end{equation}
where 
\begin{equation}
    \mathsf{c}_{1+2}^s  \equiv \frac{2}{\pi^{d}}\, \frac{ \,  \Gamma \left(\frac{d}{2}\right) \Gamma \left(\Delta_1\right) \Gamma \left(\Delta _2\right) \Gamma \left(\Delta_1+\Delta_2-\frac{d}{2}\right)}{\left(d-2 \left(\Delta_1+\Delta_2\right)\right) \Gamma \left(\Delta_1+\Delta_2\right) \Gamma \left(\Delta_1-\frac{d}{2}\right)   \Gamma \left(\Delta_2-\frac{d}{2}\right)}~.
\end{equation}

Our result in \eqref{SHE-2pf-mix} requires some explaining. A first comment is that despite looking quite different in configuration space, the correlators \eqref{General-Mixed-2pf} and \eqref{SHE-2pf-mix} show a logarithmic behavior in momentum space. In Sec.\,\ref{AdSd+1}, we found the momentum space correlators
\begin{equation}\begin{aligned}
    \langle{\cal O}_3(\vec{p}){\cal O}_3(-\vec{p})\rangle&\sim \langle{\cal O}_{1+2}(\vec{p}){\cal O}_{1+2}(-\vec{p})\rangle\sim |\vec p|^{2\Delta-d} \;,\\   
    \langle{\cal O}_3(\vec{p}){\cal O}_{1+2}(-\vec{p})\rangle&\sim \lambda_{\rm ext}\log(|\vec p|)|\vec p|^{2\Delta-d}~,
\end{aligned}
\end{equation}
and provided a unitary interpretation of the correlators by reading the $\lambda_{\rm ext}\log(|\vec p|)$ factor in the mixed correlator as an anomalous dimension expanded to first order in $\lambda_{\rm ext}$. In the same vein, in this section we have found
\begin{equation}\begin{aligned}\label{eq:mix-2}
    \langle{\cal O}^s_3(\vec{p})\widetilde {\cal O}^s_3(-\vec{p})\rangle&\sim \langle{\cal O}_{1+2}(\vec{p})\widetilde{\cal O}_{1+2}(-\vec{p})\rangle\sim |\vec p|^{0} \;,\\  
    \langle{\cal O}^s_3(\vec{p}){\cal O}_{1+2}(-\vec{p})\rangle&\sim \lambda_{\rm she}\log(|\vec p|)|\vec p|^{0}~,
\end{aligned}
    \end{equation}
which now parallels the extremal case. 
Here $\widetilde {\cal O}$ is the shadow operator of $\cal O$, i.e., it is a projector in the CFT over the conformal block of $\cal O$, see e.g. \cite{Ferrara1995,Simmons-Duffin:2012juh}.\footnote{In particular, the shadow operator $\widetilde{\cal O}$ of ${\cal O}$ satisfies 
$
\langle \widetilde{\cal O}(\vec{x}) {\cal O}(\vec{z}) \rangle\equiv \delta^d(\vec x-\vec z)$ and $\langle \widetilde{\cal O}(\vec{p}) {\cal O}(-\vec{p}) \rangle\equiv 1 = |p|^0$.}
This implies that, formally, the projector $\widetilde{\cal O}$ has many properties similar to a primary of dimension $\Delta^s=d-\Delta$, but is fundamentally different in the sense that, as a projector, it is a non-local operator and not part of the spectrum. 

At this point, a direct extrapolation of our results in Sec.\,\ref{sec:three-point-extr} would imply that a mixing between ${\cal O}_3^s$ and $\widetilde{\cal O}_{1+2}$ is taking place, but this is not possible. The reason being that, despite notation, $\widetilde{\cal O}_{1+2}$ is not formally an operator on the Hilbert space of the theory and thus it cannot be allowed to mix with proper local operators of the CFT.\footnote{We thank O. Aharony for pointing out this to us and for fruitful discussion.} It would then seem that the proper interpretation of the shadow extremal case is the one given in \cite{Aharony:2015afa} in which an anomaly is computed from \eqref{SHE-Ren-OnShell} and given the interpretation of a $\beta$-function for the composite operators. 

In a more speculative line, one could argue that our pathological results in \eqref{SHE-Ren-OnShell} are a consequence of trying to use standard tools such as inverting sources on an already non-conventional renormalized action \eqref{eq:General-Ren-OnShell-1}. If the proposed spectrum $\Delta_{\pm}$ in \eqref{Delta-pm} is consistently incorporated into the quantization of the system, then one would use its sources in \eqref{I-she-ren-0}-\eqref{phi-minus-def} rather than $\phi_3$. We expect that this should lead to a unitary CFT, thus mirroring our analysis from Sec.\,\ref{sec:three-point-extr}.

However, it is unclear how to introduce independent sources for ${\cal O}_{\pm}$ in AdS.  As it stands, the holographic dictionary prescribes clear rules on how to incorporate multi-trace sources coming from composite operators dual to fundamental fields in the bulk \cite{Witten:2001ua,Freedman:2016yue}. However, notice that our operators ${\cal O}_{\pm}$ in eq. \eqref{O+-Def} combine single- and multi-trace operators, so a more sophisticated holographic prescription would be needed to perform this ``source inversion'', at least beyond a linear analysis.

%%%%%%%%%%%%%%%%%%%%%%%%%%%%%%%%%%%%%%%%%%%%%%%%%%%%%%%%%%%%%%%%%%%%%%%%%%%%%%%%%%%%%%%%
%%%%%%%%%%%%%%%%%%%%%%%%%%%%%%%%%%%%%%%%%%%%%%%%%%%%%%%%%%%%%%%%%%%%%%%%%%%%%%%%%%%%%%%%
%%%%%%%%%%%%%%%%%%%%%%%%%%%%%%%%%%%%%%%%%%%%%%%%%%%%%%%%%%%%%%%%%%%%%%%%%%%%%%%%%%%%%%%%
\section{Summary and Discussion}\label{sec:Discussion}
%%%%%%%%%%%%%%%%%%%%%%%%%%%%%%%%%%%%%%%%%%%%%%%%%%%%%%%%%%%%%%%%%%%%%%%%%%%%%%%%%%%%%%%%
%%%%%%%%%%%%%%%%%%%%%%%%%%%%%%%%%%%%%%%%%%%%%%%%%%%%%%%%%%%%%%%%%%%%%%%%%%%%%%%%%%%%%%%%
%%%%%%%%%%%%%%%%%%%%%%%%%%%%%%%%%%%%%%%%%%%%%%%%%%%%%%%%%%%%%%%%%%%%%%%%%%%%%%%%%%%%%%%%

In this paper, we have studied an effective theory of three massive scalar fields on a fixed AdS$_{d+1}$ background in the presence of a cubic extremal coupling controlled by $\lambda_{\rm ext}$. In our bottom-up model, the conformal dimensions of each field are arranged such that
\begin{equation}\label{eq:3=1+2}
    \Delta_1+\Delta_2=\Delta_3~.
\end{equation}
The cubic interaction \eqref{OG-Action} leads to a divergent vertex, and we have shown how to construct the appropriate counterterms that render a renormalized (finite) on-shell action \eqref{eq:General-Ren-OnShell}-\eqref{eq:General-Ren-OnShell-1}. Our interpretation of the renormalized theory is that degenerate operators of the free theory are being lifted by this coupling. In particular, the new basis of primary eigenstates, to leading order in $\lambda_{\rm ext}$, is 
\begin{equation} \begin{aligned}\label{eq:definition-good-basis}
   {\cal O}_{\pm}(\vec x)&\equiv \frac{1}{\sqrt{2}}\left({\cal O}_{3}(\vec x) \pm \tilde c \, {\cal O}_{1+2}(\vec x)\right)~,
 \end{aligned}\end{equation}
 with $\tilde c$ defined in \eqref{eq:defn-ctilde}. Here ${\cal O}_i$ is the operator dual to the scalar field $\Phi_i$ and, in the jargon of AdS/CFT, single-trace, whereas ${\cal O}_{1+2}\equiv \;:\!{\cal O}_1 {\cal O}_2\!:$ is a composite (double-trace) operator. The resulting three-point functions we obtained in this new basis are 
\begin{equation}
    \langle {\cal O}_1({\vec x}){\cal O}_2({\vec y}){\cal O}_\pm({\vec z})\rangle \sim  \frac{1}{|\vec{x}-\vec{y}|^{\Delta_1+\Delta_2 - \Delta_\pm}|\vec{x}-\vec{z}|^{\Delta_1-\Delta_2 + \Delta_\pm}|\vec{y}-\vec{z}|^{-\Delta_1+\Delta_2 + \Delta_\pm}}  ~,
\end{equation}
where we have
\begin{equation}
    \Delta_\pm = \Delta_1+\Delta_2 \mp \tilde\gamma~,
\end{equation}
and $\tilde\gamma$ depends on $\lambda_{\rm ext}$ which is given in \eqref{Gen-Anom}. This expression should be understood as perturbative in $\lambda_{\rm ext}$, and here we have just computed the leading order term. Our conclusion is that extremality is an artifact of the free theory when a bulk coupling is turned on, that is, extremality is fragile from a bottom-up approach to AdS/CFT. Analogous analyses were made for the super-extremal coupling in Sec.\,\ref{sec:super-extremal}, and they are immediate extensions of the extremal ones. 

We have found some obstructions in framing the shadow-extremal interaction in the same fashion as we did in Sec.\,\ref{AdSd+1}. At this point, the only consistent interpretation is that the shadow-extremal interaction is breaking conformal invariance, along the line of the analysis in \cite{Aharony:2015afa}. If we could perform an alternative quantization procedure on ${\cal O}_{\pm}$, it might be possible to retain a CFT interpretation that would mirror the extremal case.  This claim can only be tested upon providing a more complete holographic prescription on how to consider alternative quantization of operators that are a mixture of single- and multi-trace operators.

It is worth comparing the extremal interactions in AdS with the cases where extremal correlators are persistent in the CFT. In this case, we would expect 
\begin{equation}
    \langle {\cal O}_1({\vec x}){\cal O}_2({\vec y}){\cal O}_3({\vec z})\rangle =  \frac{c_{\rm ext}}{|\vec{x}-\vec{z}|^{2\Delta_1}|\vec{y}-\vec{z}|^{2\Delta_2}}  ~,
\end{equation}
with $c_{\rm ext}$ the appropriate OPE coefficient for this extremal correlator. For a CFT with a large-$N$ limit, where the leading order contribution to correlators are Wick contractions, we can consider the redefinition
\begin{equation}\label{eq:field-re-define}
    {\cal O}_3 \to \hat{\cal O}_3 = {\cal O}_3 - c_{\rm ext} :\!{\cal O}_1 {\cal O}_2\!:+ \ldots ~,
\end{equation}
which at leading order in $N$ would give 
 \begin{equation}
    \langle {\cal O}_1({\vec x}){\cal O}_2({\vec y})\hat{\cal O}_3({\vec z})\rangle =  0 + \ldots ~.
\end{equation}
These are the field redefinitions discussed in, for example, \cite{Arutyunov:2000ima,Taylor:2007hs,Rastelli:2017udc,Rawash:2021pik}. Although \eqref{eq:definition-good-basis} and \eqref{eq:field-re-define} look very similar, the context is very different. In this quick CFT derivation, we are illustrating that one has to be careful with the choice of basis when reproducing $c_{\rm ext}$ from quantum fields on AdS$_{d+1}$, as highlighted in the literature. Our new basis \eqref{eq:definition-good-basis} is not a choice. 

There are some simple generalizations of our results. For example, one could consider extremal interactions that involve $n$-fields. It would be interesting to confirm that one reaches the same conclusions here. We only considered massive scalar fields in our analysis, and hence another generalization would be to have extremal interactions that involve fields of higher spin. It would also be interesting to inspect the fate of extremality if one goes beyond the tree-level analysis: can loops in AdS$_{d+1}$ cause corrections that affect \eqref{eq:3=1+2}? For supegravity fields on AdS$_5\times S^5$, which are cases with $\lambda_{\rm ext}=0$, a discussion on a fate of extremality at the loop level in the bulk is discussed in  \cite{Alday:2019nin,Aprile:2020uxk}.  

Another odd feature of our example is that ${\cal O}_+$ and ${\cal O}_-$ do not map to two separate fields in AdS. To further test our analysis, it would be interesting to evaluate the interacting four-point function $ \langle {\cal O}_1{\cal O}_2{\cal O}_1{\cal O}_2\rangle$.  In the CFT, we should have that both ${\cal O}_+$  and ${\cal O}_-$ can be exchanged, among other operators. In AdS the field exchanged at tree-level is only $\Phi_3$, hence it would be good to check that it is compatible with the CFT interpretation we are advocating. It would also be interesting to contrast this with the analysis of the four-point in \cite{DHoker:1998ecp, DHoker:1999mqo,Arutyunov:2000by}. In that work, an extremal boundary coupling is re-incorporated as a total derivative in the bulk (or alternative, a field redefinition), still, the analysis of the vertices should be similar.  

It would also be of interest to re-derive our results in the language of a Hamiltonian analysis AdS/CFT as in \cite{Fitzpatrick:2011hh,Fan:2011wm}. Using an operatorial dictionary and canonical quantization tools in AdS may help clarify the role of multi-trace operators for our specific setup. We expect that to leading order in $\lambda_{\rm ext}$ the chosen Hamiltonian becomes non-diagonalizable and a Jordan block emerges, indicating an anomalous dimension $\tilde \gamma$ with respect to the spectrum at $\lambda_{\rm ext}=0$. It would be interesting to check whether $\tilde \gamma$ obtained in this fashion matches our predictions in \eqref{Gen-Anom}. Although agreement between wave-function prescriptions as in \eqref{adscft} and operatorial dictionaries in AdS/CFT is expected, see for example \cite{Harlow:2011ke}, the arguments there have caveats whenever single- and multi-trace operators mix. 

Finally, we mention that in top-down examples of AdS/CFT, extremal correlators are part of the CFT data and hence it is key that the bulk couplings are zero. Still, it is somewhat mysterious why these couplings vanish from a gravitational perspective. It would be interesting to understand the rules of constructing EFT in AdS$_{d+1}$ that lead to proper CFTs. In the context of our work, is it possible to make our toy model a UV complete example of AdS/CFT?  We plan to investigate this by considering the case in \cite{Castro:2021wzn}: it is a top-down setup with an extremal bulk coupling. Since this system also has to be confronted with backreaction effects of AdS$_2$, we expect that there will be multiple competing effects, which might have an interesting interplay with extremality \cite{wip:JT}.     

\section*{Acknowledgements}
We thank Ofer Aharony, Alex Belin, Diego Correa, Juan Maldacena, Ioannis Papadimitriou, Silviu Pufu, Alessandro Tomasiello, and David Turton for helpful discussions. We thank the participants and organizers of the ``Latin American School and Workshop on Gravity and Holography'' in 2022 and the ``Amsterdam String Summer Workshop'' in 2024.  
AC has been partially supported by STFC consolidated grants ST/T000694/1 and ST/X000664/1. PJM was partially supported by CONICET and UNLP. This research was supported in part by grant NSF PHY-2309135 to the Kavli Institute for Theoretical Physics (KITP).

\appendix
\addtocontents{toc}{\protect\setcounter{tocdepth}{1}}
\numberwithin{equation}{section}

%%%%%%%%%%%%%%%%%%%%%%%%%%%%%%%%%%%%%%%%%%%%%%%%%%%%%%%%%%%%%%%%%%%%%%%%%%%%%%%%%%%%%%%%
%%%%%%%%%%%%%%%%%%%%%%%%%%%%%%%%%%%%%%%%%%%%%%%%%%%%%%%%%%%%%%%%%%%%%%%%%%%%%%%%%%%%%%%%
%%%%%%%%%%%%%%%%%%%%%%%%%%%%%%%%%%%%%%%%%%%%%%%%%%%%%%%%%%%%%%%%%%%%%%%%%%%%%%%%%%%%%%%%
\section{Examples in \texorpdfstring{AdS$_2$}{AdS2} via \texorpdfstring{$\epsilon$}{e}-prescription}\label{App:Examples}
%%%%%%%%%%%%%%%%%%%%%%%%%%%%%%%%%%%%%%%%%%%%%%%%%%%%%%%%%%%%%%%%%%%%%%%%%%%%%%%%%%%%%%%%
%%%%%%%%%%%%%%%%%%%%%%%%%%%%%%%%%%%%%%%%%%%%%%%%%%%%%%%%%%%%%%%%%%%%%%%%%%%%%%%%%%%%%%%%
%%%%%%%%%%%%%%%%%%%%%%%%%%%%%%%%%%%%%%%%%%%%%%%%%%%%%%%%%%%%%%%%%%%%%%%%%%%%%%%%%%%%%%%%

In Sec.\,\ref{AdSd+1} and Sec.\,\ref{sec:peculiar}, we have derived the renormalized correlators and actions by introducing a regulator in the problematic integrals. In this way, some of our results use analytic extensions that go beyond the formal region of convergence. In this appendix, we would like to provide a different approach to regulate these divergences. We will show how to obtain a renormalized action by using the so called ``$\epsilon$-prescription''.

In the $\epsilon$-prescription one provides boundary conditions for the fields at a finite bulk cut-off $z_0=\epsilon$ and build exact Green functions for the regularized bulk space accordingly. From there, covariant boundary terms are constructed in order to have a finite $\epsilon\to0$ limit, i.e., one implements holographic renormalization. The majority of this analysis is done in momentum space, and we will follow the procedure as done in \cite{Botta-Cantcheff:2017qir}.

Implementing this prescription at extremality for generic values of the conformal dimensions of the fields or arbitrary dimensions of the spacetime is not simple. This is mainly because the bulk integration of the vertex demands an exact result in the regularized space for an integral involving three Bessel-$K$ functions, whose result is only known for special cases; see for example \cite{Bzowski:2015yxv}. We selected concrete examples where all computations can be carried analytically. 

We will be working in AdS$_2$ and have chosen three specific combinations of matter content that lead to the different type of interactions:
\begin{equation}\nonumber
    \begin{aligned}
       & \textrm{\ref{sec:1+1=2}. {\bf Extremal}: }\qquad &&\Delta_1 = 1\,, ~ \Delta_2 = 1\,,~ \Delta_3 = 2\,, \\
       & \textrm{\ref{app:1+1+2=4}. {\bf Super-extremal}: }\qquad &&\Delta_1 = 1\,, ~ \Delta_2 = 1\,, ~\Delta_3 = 4\,, \\
        & \textrm{\ref{app:1/3+1/3+1/3=1}. {\bf Shadow-extremal}:}\qquad &&\Delta_1 = 1/3\,, ~ \Delta_2 = 1/3\,, ~\Delta_3 = 1/3\,. \\
    \end{aligned}
\end{equation}
These examples will provide an independent check for our results in Sec.\,\ref{AdSd+1} and Sec.\,\ref{sec:peculiar}.

\subsection{Example: \texorpdfstring{1$\,$+$\,$1$\,$=$\,$2}{1+1=2}}\label{sec:1+1=2}

The first example covers the case of an extremal interaction and the appropriate comparisons are with the results in Sec.\,\ref{AdSd+1}. This example is motivated by the extremal interaction found in \cite{Castro:2021wzn}, although, for the purposes of this work, the action below is just a toy model.  

We will consider two scalar fields $\Phi(x)$ and $\Psi(x)$ living on AdS$_2$, whose masses (in units of AdS length) and conformal dimensions are
\begin{equation}
    \begin{aligned}
        &\Phi(x): \quad  \quad m_\Phi^2=0 ~,\quad  \Delta_\Phi=1 ~,\\
        &\Psi(x): \quad \quad m_\Psi^2=2 ~, \quad \Delta_\Psi=2 ~. 
    \end{aligned}
\end{equation}
The effective action for these fields will be
\begin{equation}\label{App-Action}
S=S_\Phi+S_\Psi +  S_{\rm ext}~,
\end{equation}
where the first two terms are the kinetic actions for each field,
\begin{equation}
\begin{aligned}
S_\Phi&=\frac 12 \int \dd^2 x \sqrt{g}\, \partial_\mu \Phi \partial^\mu \Phi~,\\
S_\Psi&=\frac 12 \int \dd^2 x \sqrt{g}\,\left(\partial_\mu \Psi \partial^\mu \Psi +m_\Psi^2 \Psi^2\right)\;,    
\end{aligned}
\end{equation}
and we also have a cubic interaction given by 
\begin{equation}\label{eq:int-1+1=2}
 S_{\rm ext}=-\lambda_{\rm ext} \int \dd^2 x  \sqrt{g} \;\Phi^2 \Psi~.
\end{equation}
This is an extremal interaction, with coupling constant  $\lambda_{\rm ext}$. Relative to the notation in \eqref{OG-Action}-\eqref{eq:extremal-delta}, we have $\Delta_3=\Delta_\Psi$ and $\Delta_1=\Delta_2=\Delta_\Phi$.

Our conventions for the metric on AdS$_2$ follow those in Sec.\,\ref{sec:prelim}. The main difference is that now we will be placing a cutoff at a finite distance $\epsilon>0$, i.e.,
\begin{equation}
d s^2=g_{\mu\nu}\dd x^\mu \dd x^\nu=\frac{\dd z_0^2+\dd y^2}{z_0^2}~, \qquad y\in \mathbb{R}~, \qquad z_0\in[\epsilon,\infty)~.
\end{equation}
The boundary metric is defined at $z_0=\epsilon$, and in particular, we will use $\sqrt{\gamma}=\epsilon^{-1}$. The finite $\epsilon$-cutoff will render all relevant integrals regular. 
In the following, we will study the action \eqref{App-Action} perturbatively and build the renormalized on-shell action to first order in  $\lambda_{\rm ext}$. 

%%%%%%%%%%%%%%%%%%%%%%%%%%%%%%%%%%%%%%%%%%%%%%%%%%%%%%%%%%%%
\subsubsection{Equations of motion} 
%%%%%%%%%%%%%%%%%%%%%%%%%%%%%%%%%%%%%%%%%%%%%%%%%%%%%%%%%%%%
We will work mostly in Fourier space for the boundary direction $z$. To make the notation explicit, we define the Fourier transform and inverse as
\begin{equation}\label{FTransform} 
f(y,z_0)\equiv \int \frac{\dd p}{\sqrt{2\pi}} e^{i p y}\,f(p,z_0)~, \qquad f(p,z_0)\equiv \int \frac{\dd y}{\sqrt{2\pi}} e^{-i p y}\,f(y,z_0)~.
\end{equation}
It will be convenient to introduce the notation
\begin{equation}
  \partial_n:=-z_0\partial_{z_0}~, \qquad\qquad \square_p:=z_0^2(\partial_{z_0}^2-p^2)~. 
\end{equation}
With this notation, the equations of motion coming from the action \eqref{App-Action} in momentum space become
\begin{equation}\label{Phi-EOM}
\begin{aligned}    
    \square_p\Phi(p,z_0)&=2\lambda_{\rm ext} \int \frac{\dd k}{\sqrt{2\pi}} \Phi(k,z_0)\Psi(p-k,z_0)~,\\
    (\square_p-2)\Psi(p,z_0)&=\lambda_{\rm ext} \int \frac{\dd k}{\sqrt{2\pi}} \Phi(k,z_0)\Phi(p-k,z_0)~.
    \end{aligned}
\end{equation} 
The defining property of the $\epsilon$-prescription is that Dirichlet boundary conditions are
\begin{equation}\label{B-Cond}
    \Phi(p,\epsilon) = \phi(p)\;, \qquad \Psi(p,\epsilon) = \epsilon^{-1} \psi(p)\;,
\end{equation}
with no subleading pieces in $\epsilon$. This is of great help in renormalizing the action, since the map between the source and bulk field evaluated at the boundary greatly simplifies.\footnote{For general mass and bulk dimensions the boundary conditions are $\Phi_i(\vec p,\epsilon) = \epsilon^{d-\Delta_i} \Phi_i(\vec p)$.}

The equations of motion can be recursively solved by introducing a Green's function for each field. This is done as follow. For each massive field $\Phi_i$ of mass $m_i^2$ we define the Green's function
\begin{equation}
    (\square_p-m_i^2){\cal G}_i^\epsilon(p,z_0,\zeta)=+\sqrt{g}\,\delta(z_0-\zeta)\;,
\end{equation}
and 
\begin{equation}
 (\square_p-m_i^2){\cal K}_i^\epsilon(p,z_0)=0\;.
\end{equation}
The boundary conditions for these funcions are
\begin{equation}\label{eq:bcG}
\begin{aligned}
    \lim_{z_0\to\epsilon}{\cal G}_i^\epsilon(p,z_0,\zeta)=\lim_{z_0\to\infty}{\cal G}_i^\epsilon(p,z_0,\zeta)=0~, \\
      \lim_{z_0\to\epsilon}{\cal K}_i^\epsilon(p,z_0)=+1~, \qquad \lim_{z_0\to\infty}{\cal K}_i^\epsilon(p,z_0)=0 ~,   
\end{aligned}
\end{equation}
and 
\begin{equation}\label{GandK}
 \frac 1\epsilon \;\partial_n{\cal G}_i^\epsilon(p,z_0,\zeta)\big|_{z_0=\epsilon}= \sqrt{\gamma} \;\partial_n{\cal G}_i^\epsilon(p,z_0,\zeta)|_{z_0=\epsilon}\equiv {\cal K}_i^\epsilon(p,\zeta)   ~. 
\end{equation}
For $\Phi(x)$ in our example, which has $m^2_\Phi=0$, the resulting functions are
\begin{equation}\label{SolPhi1}
{\cal G}_\Phi^\epsilon(p,z_0,\zeta)=-\frac{e^{-|p| (\zeta +z_0)}}{2 |p|} \begin{cases}
e^{2 |p| z_0}-e^{2 |p| \epsilon }\,, & \epsilon\leq z_0<\zeta \\
e^{2 |p| \zeta}-e^{2 |p| \epsilon }\,, & \epsilon\leq \zeta<z_0
\end{cases}  \;,   
\end{equation}
and 
\begin{equation}
     \qquad
   {\cal K}_\Phi^\epsilon(p,z_0)=e^{-|p|(z_0-\epsilon)} ~.
\end{equation}
While for $\Psi(x)$, with $m^2_\Psi=2$, we have
\begin{equation}
{\cal G}_\Psi^\epsilon(p,z_0,\zeta)=-e^{-|p| (\zeta +z_0)}\frac{(1+ |p|\zeta) (1+|p| z_0) }{2     |p|^3 z_0 \,\zeta}\begin{cases} 
   e^{2 |p| \epsilon}\frac{ 1-|p| \epsilon}{1+|p| \epsilon}-e^{2 |p| z_0}\frac{ 1-|p| z_0}{1+|p| z_0} & z<\zeta \\
   e^{2 |p| \epsilon}\frac{ 1-|p| \epsilon}{1+|p| \epsilon}-e^{2 |p| \zeta}\frac{ 1-|p| \zeta}{1+|p| \zeta} & z>\zeta
\end{cases}  \;,
\end{equation}
and
\begin{equation}
   {\cal K}_\Psi^\epsilon(p,z_0)=e^{-|p| (z_0-\epsilon)}\frac{ (1+|p| z_0) \epsilon }{(1+|p| \epsilon)z_0}~.
\end{equation}
Then, the formal solutions to the equation of motions in \eqref{Phi-EOM} are given by
\begin{equation}\label{SolPhi}
\begin{aligned}
    \Phi(p,z_0)&={\cal K}_\Phi^\epsilon(p,z_0) \phi(p) + 2\lambda_{\rm ext} \int \frac{\dd k}{\sqrt{2\pi}} \int \dd\zeta \,\sqrt{g_\zeta} \, {\cal G}_\Phi^\epsilon(p,z_0,\zeta) \Phi(k,\zeta)\Psi(p-k,\zeta) ~,\\
    \Psi(p,z_0)&={\cal K}_\Psi^\epsilon(p,z_0) \frac{\psi(p)}{\epsilon} + \lambda_{\rm ext} \int \frac{\dd k}{\sqrt{2\pi}} \int \dd\zeta \,\sqrt{g_\zeta} \, {\cal G}_\Psi^\epsilon(p,z_0,\zeta) \Phi(k,\zeta)\Phi(p-k,\zeta) ~.
\end{aligned}
\end{equation}
These solutions are recursive in $\lambda_{\rm ext}$, but the boundary conditions \eqref{B-Cond} are met exactly at $z_0=\epsilon$, i.e., there is no $\epsilon$-expansion. 

%%%%%%%%%%%%%%%%%%%%%%%%%%%%%%%%%%%%%%%%%%%%%%%%%%%%%%%%%%%%
\subsubsection{On-Shell action} 
%%%%%%%%%%%%%%%%%%%%%%%%%%%%%%%%%%%%%%%%%%%%%%%%%%%%%%%%%%%%

One can show that in the $\epsilon$-prescription the on-shell action at leading order is
\begin{equation} \begin{aligned}\label{App:I-full}
    I \equiv I_\Phi+I_\Psi+\,I_{\rm ext}~,
 \end{aligned}\end{equation}
where we have defined
\begin{equation}\label{BoundaryI-Phi}
\begin{aligned}
    I_\Phi&=\frac {\sqrt{\gamma}}{2}   \int \dd p\, \Phi(-p,\epsilon) \, \partial_n{\cal K}_\Phi^\epsilon(p,\epsilon) \, \Phi(p,\epsilon) \,,\\
    I_\Psi&=\frac {\sqrt{\gamma}}{2}   \int \dd p\, \Psi(-p,\epsilon) \, \partial_n{\cal K}_\Psi^\epsilon(p,\epsilon) \, \Psi(p,\epsilon) \;,
\end{aligned}
\end{equation}
and $I_{\rm ext}$ is the interaction term in \eqref{eq:int-1+1=2} evaluated on-shell.
This is in contrast with the asymptotic prescription \cite{Skenderis:2002wp} where some extra terms arise due to the approximate nature of the solutions used in that prescription, see for example App. D in \cite{Bzowski:2015pba}.

%%%%%%%%%%%%%%%%%%%%%%%%%%%%%%%%%%%%%%%%%%%%%%%%%%%%%%%%%%%%%%
%%%%%%%%%%%%%%%%%%%%%%%%%%%%%%%%%%%%%%%%%%%%%%%%%%%%%%%%%%%%%%
\paragraph{Renormalization at $O(\lambda_{\rm ext}^0)$.}
%%%%%%%%%%%%%%%%%%%%%%%%%%%%%%%%%%%%%%%%%%%%%%%%%%%%%%%%%%%%%%
%%%%%%%%%%%%%%%%%%%%%%%%%%%%%%%%%%%%%%%%%%%%%%%%%%%%%%%%%%%%%%

The order zero renormalization in the $\epsilon$-prescription follows from a power expansion of the $\partial_n{\cal K}$ functions. Analytic expressions can be found for general spacetime and conformal dimensions. This is possible due to the exact relation between field and boundary data induced by \eqref{B-Cond}. 
For our case, we find that $I_\Phi$ is already finite on its own (the field is marginal) but $I_\Psi$ contains two divergent pieces that can be removed via counterterms. The renormalized action for $\lambda_{\rm ext}=0$ is then given by
\begin{equation}\label{order-0-renorm}
    \begin{aligned}
         I_{\rm ren}^0 &\equiv \lim_{\epsilon\to0} \left(I_\Phi+I_\Psi+I_{\rm ct}^0\right)\\
    &=\frac{1}{2}\int \dd p \; \phi(-p) \, |p| \,\phi(p) - \frac{1}{2}\int \dd p \; \psi(-p)  \,|p|^3\, \psi(p)\\
    &=-\frac{1}{2}\int \phi(x) \, \left(\frac{1}{\pi}\frac{1}{(x-y)^2}\right) \,\phi(y) - \frac{1}{2}\int \psi(x) \, \left(\frac{6}{\pi}\frac{1}{(x-y)^4}\right) \,\psi(y)~,
    \end{aligned}
\end{equation}
where the counterterm action reads
\begin{equation}
\begin{aligned}
    I_{\rm ct}^0
    &=-\frac{\sqrt{\gamma}}{2}\int \dd p  \,\Psi(-p,\epsilon) \Psi(p,\epsilon)-\frac{\sqrt{\gamma}}{2}\int \dd p  \,\Psi(-p,\epsilon) (\epsilon^2 p^2) \Psi(p,\epsilon)\\
    &=-\frac{1}{2}\int \dd y  \,\sqrt{\gamma} \,\Psi(y,\epsilon) \Psi(y,\epsilon)+\frac{1}{2}\int \dd y \,\sqrt{\gamma} \,\Psi(y,\epsilon) \square_\gamma \Psi(y,\epsilon)~,
\end{aligned}
\end{equation}
and we have defined $\square_\gamma:= -\epsilon^2 \partial_x^2$, the Laplacian induced at the boundary.

%%%%%%%%%%%%%%%%%%%%%%%%%%%%%%%%%%%%%%%%%%%%%%%%%%%%%%%%%%%%%%
%%%%%%%%%%%%%%%%%%%%%%%%%%%%%%%%%%%%%%%%%%%%%%%%%%%%%%%%%%%%%%
\paragraph{Renormalization at $O(\lambda_{\rm ext})$.}
%%%%%%%%%%%%%%%%%%%%%%%%%%%%%%%%%%%%%%%%%%%%%%%%%%%%%%%%%%%%%%
%%%%%%%%%%%%%%%%%%%%%%%%%%%%%%%%%%%%%%%%%%%%%%%%%%%%%%%%%%%%%%

To leading order in $\lambda_{\rm ext}$, renormalization only involves the interaction term in \eqref{App:I-full} in this prescription. We plug \eqref{SolPhi} into \eqref{eq:int-1+1=2} and expand in $\lambda_{\rm ext}$ to find the relevant divergences at $\epsilon\to0$. To order $\lambda_{\rm ext}$, we get
\begin{equation}
    \begin{aligned}\label{eq:I-ext-1+1}
           I_{\rm ext}
    &=-\lambda_{\rm ext} \int \dd z \sqrt{g}  \int \frac{\dd p \dd q }{\sqrt{2\pi}} \;\Phi(-p,z)\Phi(q,z) \Psi(p-q,z)\\
    &=-\lambda_{\rm ext} \int \dd z \sqrt{g}  \int \frac{\dd p \dd q }{\sqrt{2\pi}} \;{\cal K}_\Phi^\epsilon(p,z) \phi(-p)\, {\cal K}_\Phi^\epsilon(q,z) \phi(q)\, {\cal K}_\Psi^\epsilon(p-q,z) \frac{\psi(p-q)}{\epsilon}\\
    &=-\lambda_{\rm ext} \int \frac{\dd p \dd q \dd k}{\sqrt{2\pi}} \delta(p+q+k)\;\phi(p)\,\phi(q) \,\frac{\psi(k)}{\epsilon} \, {\cal I}(p,q,k)~,
    \end{aligned}
\end{equation}
where we introduce the $\delta(p+q+k)$ in order to define more generally the integral
\begin{equation}
   {\cal I}(p,q,k):= \int_\epsilon^\infty \dd z\,\sqrt{g} \, {\cal K}_\Phi^\epsilon(p,z) {\cal K}_\Phi^\epsilon(q,z) {\cal K}_\Psi^\epsilon(k,z)~.
\end{equation}
For the values of masses (conformal dimensions) chosen here, we can evaluate and analyze this integral, although it is worth mentioning that it is not straightforward for generic fields; see, e.g., \cite{Bzowski:2015pba}. Solving this integral explicitly gives
\begin{equation}
    \begin{aligned}
        \label{renorm-lambda1}
     {\cal I}(p,q,k)&=\frac{1}{2
   (1+|k| \epsilon )}\Bigg[\epsilon^{-1}+(|k|-|p|-|q|)-\epsilon\; e^{(|k|+|p|+|q|) \epsilon }
   (|k|+|p|+|q|)^2  \\
    &\qquad   \times \left( \text{Ei}(-((|k|+|p|+|q|)
   \epsilon ))+2 |k|\frac{ \Gamma (0,(|k|+|p|+|q|) \epsilon )}{(|k|+|p|+|q|)}\right)\Bigg]\\
   &= \frac{1}{2 \epsilon } -\frac{|p|+|q|}{2}+\frac{\epsilon}{2} |k|   (|p|+|q|)  \\
   &\qquad+\frac{\epsilon}{2}  (|k|^2-(|q|+|p|)^2) \ln   \Big(\epsilon(|k|+|p|+|q|)e^{\gamma} \Big) +O(\epsilon^2)~,
    \end{aligned}
\end{equation}
where in the second line we expanded the answer in powers of $\epsilon$ and only kept the terms that do not vanish in the $\epsilon\to0$ limit of the on-shell action \eqref{eq:I-ext-1+1}. Our final task for this exercise is to understand the nature of each of these terms and how to construct the appropriate counterterms that will render a finite on-shell action. 

The first two terms in \eqref{renorm-lambda1} are simple to analyze. The first term is divergent when replaced in \eqref{eq:I-ext-1+1}. The counterterm that we need to cancel this divergence can be found by replacing it back into the on-shell action \eqref{eq:I-ext-1+1} and casting it covariantly. We find 
\begin{equation}
    \begin{aligned}
        \label{Ict1}
     I_{\rm ct;1}&\equiv - \lambda_{\rm ext} \int \frac{\dd p \dd q \dd k}{\sqrt{2\pi}} \delta(p+q+k)\;\phi(p)\,\phi(q) \,\frac{\psi(k)}{\epsilon}\left(\frac{1}{2 \epsilon }\right) \\
    &=-\frac{\lambda_{\rm ext}}{2} \int \dd y \sqrt{\gamma}\;\Phi(y,\epsilon)\Phi(y,\epsilon) \Psi(y,\epsilon)~.
    \end{aligned}
\end{equation}
The second term in \eqref{renorm-lambda1} also requires the introduction of a counterterm. This is found again by casting this divergence covariantly, which gives
\begin{equation}
    \begin{aligned}
        \label{Ict2}
I_{\rm ct;2}&\equiv -\lambda_{\rm ext} \int \frac{\dd p \dd q \dd k}{\sqrt{2\pi}} \delta(p+q+k)\;\phi(p)\,\phi(q) \,\frac{\psi(k)}{\epsilon}\left(-\frac{|p|+|q|}{2}\right)\\
&=  \lambda_{\rm ext} \int\dd y\sqrt{\gamma}\;\Phi(y,\epsilon) \Psi(y,\epsilon)\;\Pi_{\Phi}(y,\epsilon)+O(\lambda_{\rm ext}^2)~.
    \end{aligned}
\end{equation}
In $I_{\rm ct;2}$ we see that the appearance of $|p|$ forced us to use the conjugated momentum $\Pi_{\Phi}$ to build the correct counterterms, where the conjugated momentum is defined in  \eqref{def-Gen-Pi}. Notice that the use of $\Pi_{\Phi}$ counterterms induces corrections to higher orders in $\lambda_{\rm ext}$ which we made explicit in \eqref{Ict2}. Up to this stage then we have
\begin{equation}
\begin{aligned}
    \lim_{\epsilon\to0}\left( I_{\rm ext}+ \,I_{\rm ct;1} + I_{\rm ct;2} \right) = 
-\lambda_{\rm ext} \int \frac{\dd p \dd q \dd k}{\sqrt{2\pi}} \delta(p+q+k)\;\phi(p)\,\phi(q) \,\frac{\psi(k)}{\epsilon} \, \hat{\cal I}(p,q,k)~,
\end{aligned}
\end{equation}
where 
\begin{equation}\label{eq:hat-I-defn}
\begin{aligned}
     \hat{\cal I}(p,q,k) :=
   &\frac{ \epsilon}{2}   (|q|+|p|)|k|+\frac{\epsilon}{2}  (|k|^2-(|q|+|p|)^2) \ln\left((|k|+|q|+|p|)\epsilon e^{\gamma}\right)~.
\end{aligned}
\end{equation}
These are the contributions that will lead to the renormalized correlator. Each term in $\hat{\cal I}(p,q,k)$ can be treated independently. The first piece is not difficult to transform cast back in position space,
\begin{equation} \begin{aligned}
    &-\lambda_{\rm ext} \int \frac{\dd p \dd q \dd k}{\sqrt{2\pi}} \delta(p+q+k)\;\phi(p)\,\phi(q) \,\psi(k) \frac{(|q|+|p|)|k|}{2}\\
    &\quad=-\frac{\lambda_{\rm ext}}{2} \int  \dd x\dd y \dd z  \;\phi(x)\,\phi(y) \,\psi(z) \left(\frac {1}{\pi^2}\frac{1}{(y-z)^2(x-y)^2}+\frac {1}{\pi^2}\frac{1}{(x-z)^2(x-y)^2}\right)\\
    &\quad=-\frac{\lambda_{\rm ext}}{2} \int  dx\dd y \dd z  \;\phi(x)\,\phi(y) \,\psi(z) \left(\frac{1}{\pi^2}\frac {a^2+b^2}{ a^2 b^2 (a-b)^2}\right)~,\nn
 \end{aligned}\end{equation} 
where we defined $a:=(x-z)$ and $b:=(y-z)$. For the second term in \eqref{eq:hat-I-defn}, it is convenient to use the identity
\begin{equation}
   (|k|+|q|+|p|) \ln\left((|k|+|q|+|p|)\epsilon e^{\gamma}\right)=\partial_{\alpha} \Big( (|k|+|q|+|p|)^\alpha (\epsilon e^{\gamma})^{\alpha-1} \Big)_{\alpha=1}~.
\end{equation}
With this we then have 
\begin{equation} \begin{aligned}\label{eq:corcho-FT}
&\frac{\lambda_{\rm ext}}{2} \int\!\! \frac{\dd p \dd q \dd k}{\sqrt{2\pi}} \delta(p+q+k)\;\phi(p)\,\phi(q) \,\psi(k)   (|q|+|p|-|k|)  \partial_{\alpha} \Big( (|q|+|p|+|k|)^\alpha (\epsilon e^{\gamma})^{\alpha-1} \Big)_{\alpha=1}\\
&= \frac{\lambda_{\rm ext}}{2} \int \dd x\dd y \dd z  \; \phi(x)\,\phi(y) \,\psi(z) \,\partial_{\alpha} \Big(  (\epsilon e^{\gamma})^{\alpha-1}    \;FT_\alpha  \Big)_{\alpha=1}\\
   &= \frac{\lambda_{\rm ext}}{2} \int \dd x\dd y \dd z  \; \phi(x)\,\phi(y) \,\psi(z) \, \left(\frac{1}{\pi^2}\frac {a^2+b^2}{ a^2 b^2 (a-b)^2}+\frac{1}{\pi^2}\frac { \ln \left(\frac{4 e \epsilon ^2 (a-b)^2}{a^2 b^2}\right)}{ a^2 b^2 }\right)~.
 \end{aligned}\end{equation}
The details of the Fourier transform are explained in App.\,\ref{app:conventions}, and the definition of $FT_\alpha$ is in \eqref{eq:defn-FT}. By adding the two expressions above we get
\begin{equation}\begin{aligned}\label{A+B}
    \lim_{\epsilon\to0}\left( I_{\rm ext}+ \,I_{\rm ct;1} + I_{\rm ct;2} \right) =& \frac{\lambda_{\rm ext}}{2\pi^2} \int \dd x\dd y \dd z  \; \phi(x)\,\phi(y) \,\psi(z) \\ & \quad \qquad\qquad\times\left( \frac { \ln \left(\frac{(x-y)^2}{(x-z)^2 (y-z)^2}\right)}{(x-z)^2 (y-z)^2 }+\frac { \ln \left(4 e \epsilon ^2\right)}{(x-z)^2 (y-z)^2 }\right)~.
\end{aligned}\end{equation} 
The logarithmic divergence here can be removed by the counterterm
\begin{equation} \begin{aligned}\label{Ict3}
    I_{\rm ct;3}
    &= -\frac{\lambda_{\rm ext}}{2} \ln \left(4 e \epsilon ^2\right) \int \sqrt{\gamma} \; \Pi_\Phi(x)\,\Pi_\Phi(x)\,\Psi(x)+O(\lambda_{\rm ext}^2)~.
 \end{aligned}\end{equation}
The counterterms at order $\lambda_{\rm ext}$ are then obtained by combining \eqref{Ict1}, \eqref{Ict2} and \eqref{Ict3}. The renormalized action we get is
\begin{equation} \begin{aligned}\label{Ifin-example}
   I_{\rm ren}&= I_{\rm ren}^0 + \lim_{\epsilon\to0}\left( I_{\rm ext}+ \,I_{\rm ct;1} + I_{\rm ct;2} + I_{\rm ct;3}\right)\\
   &=-\frac{1}{2\pi}\int \dd x_1 \dd x_2 \left( \phi(x_1) \, \frac{1}{(x_1-x_2)^2} \,\phi(x_2) +\psi(x_1) \, \frac{6}{(x_1-x_2)^4} \,\psi(x_2)\right) \\
   &\qquad+ \frac{\lambda_{\rm ext}}{2\pi^2} \int \dd x_1 \dd x_2 \dd x_3 \; \phi(x_1)\,\phi(x_2) \,\psi(x_3) \frac { \ln \left(\frac{ (x_1-x_2)^2}{(x_1-x_3)^2 (x_2-x_3)^2}\right)}{ (x_1-x_3)^2 (x_2-x_3)^2 }+ O(\lambda_{\rm ext }^2)
 \end{aligned}\end{equation}
 This agrees perfectly with the renormalized action in \eqref{eq:General-Ren-OnShell}-\eqref{eq:General-Ren-OnShell-1}. 

%%%%%%%%%%%%%%%%%%%%%%%%%%%%%%%%%%%%%%%%
%%%%%%%%%%%%%%%%%%%%%%%%%%%%%%%%%%%%%%%%
\subsection{Example: \texorpdfstring{1$\,$+$\,$1$\,$+$\,$2$\,$=$\,$4}{1+1+2=4}}\label{app:1+1+2=4}
%%%%%%%%%%%%%%%%%%%%%%%%%%%%%%%%%%%%%%%%
%%%%%%%%%%%%%%%%%%%%%%%%%%%%%%%%%%%%%%%%

Our second example involves again two fields in AdS$_2$, which we denote as $\Phi(x)$ and $\Psi(x)$, but now  the  conformal dimensions are
\begin{equation}
    \Delta_\Phi = 1 ~, \qquad \Delta_\Psi=4~.
\end{equation}
In addition to the kinetic and mass terms, we will introduce the following cubic interaction 
\begin{equation}\label{eq:int-1+1+2=4}
 S_{\rm se}=-\lambda_{\rm se} \int \dd^2 x  \sqrt{g} \;\Phi^2 \Psi~.
\end{equation}
This corresponds to a super-extremal interaction, since $\Delta_\Psi= \Delta_\Phi + \Delta_\Phi +2$. In the notation of Sec.\,\ref{sec:super-extremal}, we have $\Delta_1=\Delta_2=1$ and $\Delta_3=4$.

To construct the renormalized on-shell action, the steps follow the exact same logic as the extremal case in App.\,\ref{sec:1+1=2}. The main difference is that now $\Psi(x)$ has a different value of its mass: $m_\Psi^2=4(4-1)=12$, in AdS units. 
The order $\lambda_{\rm se}^0$ piece of the action follows from \eqref{order-0-renorm}; the interaction term follows very closely the steps starting from \eqref{eq:I-ext-1+1}, with the only difference being that we need extra counterterms. The resulting renormalized on-shell action, to leading order in $\lambda_{\rm se}$, is
\begin{equation} \begin{aligned}
   I_{\rm ren}=& \lim_{\epsilon\to0}\left( I+I_{\rm ct}^0+I_{\rm ct}^1\right)\\
   &-\frac{1}{2}\int \dd x_1 \dd x_2 \left( \phi(x_1) \, \left(\frac{1}{\pi}\frac{1}{(x_1-x_2)^2}\right) \,\phi(x_2) +\psi(x_1) \, \left(\frac{112}{5\pi}\frac{1}{(x_1-x_2)^8}\right) \,\psi(x_2)\right) \nn\\
   &+ \frac{\lambda_{\rm se}}{5 \pi ^2} \int\!\! \dd x_1 \dd x_2 \dd x_3 \, \phi(x_1)\,\phi(x_2) \,\psi(x_3) 
   \frac {  (x_1-x_2)^2}{ (x_1-x_3)^4 (x_2-x_3)^4 }\ln \left(\frac{ (x_1-x_2)^2}{(x_1-x_3)^2 (x_2-x_3)^2}\right)
 \end{aligned}\end{equation}
and the anomalous counterterms used to regulate the super-extremal interaction are
\begin{equation} \begin{aligned}
    I_{\rm ct}^1 =&-\frac{\lambda_{\rm se}\ln(
    \epsilon^2 )}{48} \int \dd x \sqrt{\gamma}\;\Phi^2 \,(\square_\gamma^2\Psi) +\frac{7\lambda_{\rm se}\ln(
    \epsilon^2 )}{60} \int \dd x \sqrt{\gamma}\;(\square_\gamma\Psi)\left(\Pi_\Phi^2+\Phi(\square_\gamma\Phi)\right) \\
    &-\frac{3\lambda_{\rm se}\ln(\epsilon^2 )}{40} \int \dd x \sqrt{\gamma}\;\Psi\left((\square_\gamma\Phi)^2+\Phi(\square_\gamma^2\Phi)\right)
    +\frac{3\lambda_{\rm se}\ln(\epsilon^2 )}{10} \int \dd x \sqrt{\gamma}\;\Psi\Pi_\Phi(\square_\gamma\Pi_\Phi)~.\nn
 \end{aligned}\end{equation}
This agrees with our findings in Sec.\,\ref{sec:super-extremal}, up to total derivatives.
 
%%%%%%%%%%%%%%%%%%%%%%%%%%%%%%%%%%%%%%%%
%%%%%%%%%%%%%%%%%%%%%%%%%%%%%%%%%%%%%%%%
\subsection{Example: \texorpdfstring{1/3$\,$+$\,$1/3$\,$+$\,$1/3$\,$=$\,$1}{1/3+1/3+1/3=1}}\label{app:1/3+1/3+1/3=1}
%%%%%%%%%%%%%%%%%%%%%%%%%%%%%%%%%%%%%%%%
%%%%%%%%%%%%%%%%%%%%%%%%%%%%%%%%%%%%%%%%

Finally, we consider an EFT in AdS$_2$ of three scalar fields $\Phi_i$ with equal mass which we set to be  
\begin{equation}\label{eq:mass-ex-shadow}
    m^2_i = \Delta_i(\Delta_i-1)=-\frac{2}{9}~.
\end{equation}
In addition to the kinetic and mass terms for each field, we will have the bulk interaction
\begin{equation}\label{eq:int-shadow-app}
    S_{\rm int}=-\lambda_{\rm she} \int \dd^2 x \sqrt{g}\, \Phi_1\Phi_2\Phi_3~.
\end{equation}
Due to the specific mass selected in \eqref{eq:mass-ex-shadow}, this interaction term is interesting. If we decide to quantize all the fields as operators with 
\begin{equation}
 \Delta_i^s =\frac{1}{3}~,
\end{equation}
then the interaction \eqref{eq:int-shadow-app} is {\it shadow-extremal}, where the sum of conformal dimensions adds up to the number of dimensions of the CFT. However, if we quantize all fields as operators  with 
\begin{equation}
 \Delta_i = 1-\Delta_i^s = \frac{2}{3}~,
\end{equation}
then the interaction has no peculiar features, it is just a vanilla interaction in AdS with no explicit singularities. 

Given this, our strategy is to quantize the field such that the dual operator has $\Delta_i$ (corresponding to Dirichlet boundary conditions) and obtain a renormalized action for the interacting theory at leading order in $\lambda_{\rm she}$. This is the standard quantization of the fields. To obtain the alternative quantization of $\Phi_i$, we will then do an appropriate Legendre transform on this renormalized action, which will lead to a treatment of the interaction when the field is dual to the operator with $\Delta_i^s$.  An existing example in the literature that follows this strategy in AdS$_4$ is described in \cite{Freedman:2016yue}; however, they consider $\lambda_{\rm she}=0$.

In the standard quantization, the procedure to renormalize the theory is very standard, and therefore we will omit details. The only point to highlight is that due to the spectrum degeneracy, a finite bulk counterterm can be written. The existence of finite bulk counterterms is common to all the peculiar couplings studied in this work and has already manifested in, e.g., \eqref{CT-constant-shift-1}. In this example, for the standard quantization, the boundary term is 
\begin{equation}\begin{aligned}\label{eq:bndy-shadow-c}
     I_{\rm bndy}&= c \int\!\!\dd x\, \sqrt{\gamma}\,\Phi_1\Phi_2\Phi_3\\
    &= c \int \!\! \dd x\, \phi_1(x)\phi_2(x)\phi_3(x)~.
\end{aligned}
\end{equation} 
Here $c$ is a constant, and in the second line, we have evaluated this term on-shell, where $\phi(x)$ is the source in standard quantization. This boundary term will lead to three-point functions in either quantization. In comparison, the analysis of \cite{Freedman:2016yue} considers only the effect of this type of boundary terms.

Including the boundary and bulk interaction terms, the on-shell renormalized action in the standard quantization is
\begin{equation} \begin{aligned}
I_{\rm ren}
=&-\frac{1}{2}\int  \, \prod_{i=1}^3 \, \left(\frac{\Gamma \left(\frac{2}{3}\right)}{3 \sqrt{\pi }\, \Gamma
\left(\frac{1}{6}\right)}\frac{\phi_i (x_1)\phi_i (x_2)}{|x_1-x_2|^{4/3}}\right) \, + c \int \!\!  \phi_1 (x)\phi_2 (x)\phi_3 (x)  \\
& -\lambda_{\rm she} \int \; \left( \frac{ \Gamma
\left(\frac{1}{3}\right)^3}{2 \sqrt{\pi} \Gamma \left(\frac{1}{6}\right)^3 } \frac { \phi_1 (x_1)\,\phi_2 (x_2) \,\phi_3 (x_3)}{ |x_1-x_2|^{2/3}|x_1-x_3|^{2/3}|x_2-x_3|^{2/3} }\right)+ O(\lambda^2_{\rm she}) ~.
 \end{aligned}\end{equation}

Next, we would like to perform the Legendre transform that switches the quantization to be $\Delta_i ^s$. This is done, as discussed in Sec.\,\ref{sec:shadow-extremal}, by defining $\phi^s_i$, the source of ${\cal O}^s_i$ as
\begin{equation}\label{eq:phi-minus-def}
    \phi^s_{i}(\vec x)\equiv - \frac{\delta I_{\rm ren}[\phi_{i}]}{\delta\phi_{i}(\vec x)}~. 
\end{equation}
This relation is non-linear and therefore difficult to cast  $\phi_i $ in terms of $\phi_i ^s$. This can be done recursively as an expansion in the coupling and sources, which leads to
\begin{equation} \begin{aligned}\label{eq:phi+-phi-}
    \phi_i (z)=&\int \dd x \left(\frac{3 \Gamma
   \left(\frac{1}{3}\right)}{\sqrt{\pi } \Gamma\left(-\frac{1}{6}\right)} \frac{1}{|x-z|^{2/3}}\right)\phi_i ^s(x)+ O(\lambda_{\rm she})+O((\phi^s)^2) \\
   =&\int \dd x \left(\frac{3 \Gamma
   \left(\frac{1}{3}\right)}{\sqrt{\pi } \Gamma\left(-\frac{1}{6}\right)} \frac{1}{|x-z|^{2/3}}\right)\phi_i ^s(x) \\
   & -\frac{\lambda_{\rm she}}{4} \int \dd x_1 \dd x_2 \left(\frac{\Gamma \left(\frac{1}{6}\right)^3}{2
   \pi  \Gamma \left(\frac{5}{6}\right)^3}\right) \left( \frac { \ln\left(\frac{|x_1-x_2|^2 \tilde\epsilon^2}{|x_1-z|^2|x_2-z|^2} \right) \epsilon_{ijk}\phi_j ^s(x_1)\,\phi_k ^s(x_2)}{ |x_1-z|^{1/3}|x_2-z|^{1/3} |x_1-x_2|^{1/3}}\right) \\
   & - \lambda_{\rm she} \int \dd x_1  
   \left(\frac{  \Gamma
   \left(-\frac{2}{3}\right) \Gamma
   \left(\frac{1}{6}\right) }{2^{2/3} \Gamma
   \left(\frac{5}{6}\right)^3  } \right)
   \left( \frac{1}{ | x_1-z| }+\ln(\tilde\epsilon^2)\delta(x_1-z)\right)\times\\
&\qquad\qquad\qquad \qquad\qquad\qquad \qquad\qquad\qquad \times   
\epsilon_{ijk}\,\phi_{j+k} ^s(x_1)
   \\
   &+ O(\lambda_{\rm she}^2) +O((\phi^s)^3) ~.
 \end{aligned}\end{equation}
In obtaining this result, a few key steps need to be highlighted. The inversion process of the source demanded performing integrals very similar to those in \eqref{Inverse-Int-SHE}, however with one of the powers in the denominators on the right-hand side adding up to zero. This makes the integral divergent, and we regulated via dimensional regularization: 
 the parameter $\tilde\epsilon$ in \eqref{eq:phi+-phi-} is that regulator. The constant $c$ does not appear above explicitly, since its appearance can be reabsorbed in the definition of the $\tilde\epsilon$ regulator.  This is because \eqref{eq:bndy-shadow-c} plays an analogous role as \eqref{CT-constant-shift-1} in the extremal case; that is, $c$ can be used to remove all dependence in $\tilde\epsilon$ on the finite on-shell action. Notice that the fourth line above is a non-trivial contact contribution to the third line of the same equation, much like \eqref{eq:Iren-c}. In Sec.\,\ref{sec:shadow-extremal}, we have adjusted the boundary counterterm such that $\tilde \epsilon=1$. Finally, we have used the anti-symmetric $\epsilon_{ijk}$ Levi-Civita symbol to shorten notation.
We get for the shadow-extremal on-shell action
\begin{equation} \begin{aligned}\label{I-she-example}
I^{\rm she}_{\rm ren}
=&-\frac{1}{2}\int \dd x_1 \dd x_2 \, \prod_{i=1}^3 \, \left(\frac{3 \Gamma
   \left(\frac{1}{3}\right)}{\sqrt{\pi } \Gamma\left(-\frac{1}{6}\right)} \frac{\phi_i ^s(x_1)\phi_i ^s(x_2)}{|x_1-x_2|^{2/3}}\right)\nn \\
& +\lambda_{\rm she} \int \dd x_1 \dd x_2 \dd x_3 \left(\frac{\Gamma \left(\frac{1}{6}\right)^3}{2
   \pi  \Gamma \left(\frac{5}{6}\right)^3}\right) \left( \frac { \ln\left(\frac{|x_1-x_2|^2 
   \tilde\epsilon^2
   }{|x_1-x_3|^2|x_2-x_3|^2} \right) \phi_1 ^s(x_1)\,\phi_2 ^s(x_2)\,\phi_3^s(x_3)}{ |x_1-x_3|^{1/3}|x_2-x_3|^{1/3} |x_1-x_2|^{1/3}}\right)\nn \\
   & - \lambda_{\rm she} \int \dd x_1 \dd x_2 \left(\frac{  \Gamma
   \left(-\frac{2}{3}\right) \Gamma
   \left(\frac{1}{6}\right) }{2^{2/3} \Gamma
   \left(\frac{5}{6}\right)^3  } \right) 
   \left( \frac{1}{ | x_1-x_2| }+\ln(\tilde\epsilon^2)\delta(x_1-x_2)\right)\nn \\
&\qquad\qquad\qquad \qquad\qquad\qquad \qquad\qquad\qquad \times\left(\phi_{1+2} ^s(x_1)\,\phi_3^s(x_2)+\text{ perm. }\right)\nn\\
   &+ O(\lambda_{\rm she}^2) +O((\phi^s)^4) \nn\;.
 \end{aligned}\end{equation}
The interpretation now follows exactly as in Sec.\,\ref{sec:shadow-extremal}. 
However, it is important to note that the coefficients here are not the ones appearing in \eqref{SHE-Ren-OnShell}. This is because the coefficients in front of the correlators, in our normalization of $\phi^s$, depend on how many of the operators are in the upper and lower branch, i.e. how many inversions $\phi\to\phi^s$ where needed. We needed to invert all the sources in this example, while in the toy example in Sec.\,\ref{sec:shadow-extremal} we did only a single inversion.

%%%%%%%%%%%%%%%%%%%%%%%%%%%%%%%%%%%%%%%%%%%%%%%%%%%%%%%%%%%%%%%%%%%%%%%%%%%%%%%%%%%%%%%%
%%%%%%%%%%%%%%%%%%%%%%%%%%%%%%%%%%%%%%%%%%%%%%%%%%%%%%%%%%%%%%%%%%%%%%%%%%%%%%%%%%%%%%%%
%%%%%%%%%%%%%%%%%%%%%%%%%%%%%%%%%%%%%%%%%%%%%%%%%%%%%%%%%%%%%%%%%%%%%%%%%%%%%%%%%%%%%%%%
\section{Integral identities}\label{app:conventions}
%%%%%%%%%%%%%%%%%%%%%%%%%%%%%%%%%%%%%%%%%%%%%%%%%%%%%%%%%%%%%%%%%%%%%%%%%%%%%%%%%%%%%%%%
%%%%%%%%%%%%%%%%%%%%%%%%%%%%%%%%%%%%%%%%%%%%%%%%%%%%%%%%%%%%%%%%%%%%%%%%%%%%%%%%%%%%%%%%
%%%%%%%%%%%%%%%%%%%%%%%%%%%%%%%%%%%%%%%%%%%%%%%%%%%%%%%%%%%%%%%%%%%%%%%%%%%%%%%%%%%%%%%%

In this appendix, we collect some useful integrals used throughout the text. 

\paragraph{Extremal interactions in Sec.\,\ref{AdSd+1}.} The first integral is the Feynman parametrization technique, which reads 
\begin{equation}\label{Feyn}
\frac{1}{X^a Y^b}=\frac{\Gamma(a+b)}{\Gamma(a)\Gamma(b)}\int_0^1 \dd u\frac{u^{a-1}(1-u)^{b-1}}{\left(u X + (1-u)Y\right)^{a+b}}~.
\end{equation}
Here $\Re(a)>0$ and $\Re(b)>0$. Another integral we used is
\begin{equation}\label{RInt}
\int \frac{d^d z}{\left(\vec z^2+2 \vec z \cdot \vec x+X^2\right)^a}
=\frac{\Gamma(a-d/2)}{\Gamma(a)}\frac{\pi^{d/2}}{\left(X^2-\vec x^2\right)^{a-\frac d2}}~,
\end{equation}
where the integral is over $\mathbb{R}^d$ and $a>0$. 

In Sec.\,\ref{sec:three-point-extr} we introduced an IR regulator to tame a divergent integral. In that context, we used the following identity
\begin{equation}
\begin{aligned}\label{Aux-Ext1}
\int_{0}^{1/\epsilon} \dd w'_0  \frac{(w'_0)^{2b-1} }{[(w'_0)^2+ X^2]^{b}}=\frac{1}{b}\left(\epsilon^2 X^2\right)^{-b}\, _2F_1\left(b,b;b+1;-\frac{1}{X^2\epsilon^2}\right)~,
\end{aligned}    
\end{equation}
where $\Re(b)>0$ and $\Re(X^2)>0$. In the limit $\epsilon \to 0$ we have
\begin{equation}\label{Aux-Ext2}
   _2F_1\left(b,b;b+1;-\frac{1}{X^2\epsilon^2}\right) = -b (M\epsilon)^{2b}\left(\log
   \left( X^2 \epsilon ^2 \right)  +\psi(b) + \gamma    + O(\epsilon^2) \right)~, 
\end{equation}
which is the divergent piece reported in \eqref{eq:IR-reg-int}. To get to \eqref{eq:R-epsilon-final}, we also used
\begin{equation}
    \int_0^1\dd u \,u^{b-1}(1-u)^{c-1} = \frac{\Gamma(b)\Gamma(c)}{\Gamma(b+c)}~.
\end{equation}

\paragraph{Shadow-extremal interactions in Sec.\,\ref{sec:shadow-extremal}.}
Here we needed
\begin{equation}\label{Inverse-1}
    \int \dd^d w \frac{1}{|\vec x - \vec w|^{2\Delta}} \frac{1}{|\vec w - \vec y|^{2(d-\Delta)}}=\pi^{d} \frac{\Gamma(\Delta-d/2)\Gamma(d/2-\Delta)}{\Gamma(\Delta)\Gamma(d-\Delta)} \delta^d(\vec x - \vec y)~.
\end{equation}
We also used
\begin{equation}\label{Inverse-Free2-SHE}
    \int \dd^d z\frac{\ln(|\vec{x}-\vec{z}|^2)}{ |\vec{x}-\vec{z}|^{2(d-\Delta)}}\frac{1}{|\vec z - \vec w|^{2\Delta}}=-\pi ^{\frac d2}\frac{ \Gamma
   \left(\frac{d}{2}\right) \Gamma
   \left(\frac{d}{2}-\Delta\right)
   \Gamma \left(\Delta-\frac{d}{2}\right)}{\Gamma
   \left(\Delta\right) \Gamma
   \left(d-\Delta \right)}\frac{1}{|\vec x- \vec w|^{d}}~,
\end{equation}
And finally, we  have
\begin{equation} \begin{aligned}\label{Inverse-Int-SHE}
& \int \dd^d z 
\frac{1}{|\vec{x}_1-\vec{z}|^{2 \Delta_1}|\vec{x}_2-\vec{z}|^{2 \Delta_2}|\vec x_3 - \vec z|^{2\Delta_3}}\\
&\qquad =\pi^{\frac d2}    \prod_{i=1}^3\frac{ \Gamma\left(\frac{d}{2}-\Delta_i\right)}{\Gamma\left(\Delta_i\right)}
   \frac{1}{|\vec{x}_1-\vec{x}_2|^{d-2 \Delta_3}|\vec{x}_1-\vec{x}_3|^{d-2 \Delta_2}|\vec{x}_2-\vec{x}_3|^{d-2\Delta_1}}~,
 \end{aligned}\end{equation}
whenever $\Delta_1+\Delta_2+\Delta_3=d$.

%%%%%%%%%%%%%%%%%%%%%%%%%%%%%%%%%%%%%%%%
\paragraph{Fourier Transforms in App.\,\ref{sec:1+1=2}.}
%%%%%%%%%%%%%%%%%%%%%%%%%%%%%%%%%%%%%%%%

In \eqref{eq:corcho-FT} we introduced $FT_\alpha$ which is defined as
\begin{equation} \begin{aligned}\label{eq:defn-FT}
    FT_\alpha&= \int \frac{\dd p \dd q \dd k}{(2\pi)^2} e^{-i p x - i q y- i k z }\delta(p+q+k)\;  (|q|+|p|-|k|) (|q|+|p|+|k|)^\alpha ~.
 \end{aligned}\end{equation}
To solve this integral we introduce polar coordinates in $p=l\cos\theta$ and $k=l\sin\theta$ and define $a=x-z$ and $b=y-z$, which  gives   
    \begin{equation} \begin{aligned}
  FT_\alpha  &=\int_{-\pi}^\pi \frac{ \dd\theta }{(2\pi)^2} \;  (|\cos\theta|+|\sin\theta|-|\cos\theta+\sin\theta|) \\
    &\qquad\qquad\qquad\qquad  \times(|\cos\theta|+|\sin\theta|+|\cos\theta+\sin\theta|)^\alpha\int_0^\infty \,\dd l\,  e^{-i l( a\cos\theta + b\sin\theta) }\;l^{\alpha+2}\\
    &= \sin \left(\frac{\pi  \alpha
   }{2}\right) \frac{\Gamma (\alpha
   +3)}{2 \pi^2} \int_{0}^\pi  \dd\theta  \;  \frac{|\cos\theta|+|\sin\theta|-|\cos\theta+\sin\theta| }{| a\cos\theta + b\sin\theta| ^{\alpha +3}}\\
    &\qquad\qquad\qquad\qquad\qquad\qquad\qquad\qquad\qquad  \times(|\cos\theta|+|\sin\theta|+|\cos\theta+\sin\theta|)^\alpha\nn\\
   &= \sin \left(\frac{\pi  \alpha
   }{2}\right) \frac{\Gamma (\alpha
   +3)}{2 \pi^2} \left( \int_{\pi/2}^{3\pi/4} \dd\theta   \frac{(-2\cos\theta)(2\sin\theta)^\alpha}{| a\cos\theta + b\sin\theta| ^{\alpha +3}} + \int_{3\pi/4}^\pi \dd\theta   \frac{(2\sin\theta)(-2\cos\theta)^\alpha}{| a\cos\theta + b\sin\theta| ^{\alpha +3}}\right) ~.
 \end{aligned}\end{equation}
The angular integrals can be solved and, for example, the first one gives
\begin{equation} \begin{aligned}\label{final-angle}
   \int_{\pi/2}^{3\pi/4} \dd\theta  \;  \frac{(-2\cos\theta)(2\sin\theta)^\alpha}{| a\cos\theta + b\sin\theta| ^{\alpha +3}}&=2^{\alpha +1} \int_{0}^{\pi/4}\dd\varphi \, \frac{\sin\varphi \cos ^{\alpha }\varphi }{ | b \cos \varphi -a \sin \varphi| ^{\alpha+3} }\\
    &=\frac{2^{\alpha +1}}{ (\alpha +1) (\alpha +2)}
    \left( \frac{1}{b^{\alpha+1} a^2 } + \frac{b-a (\alpha +2)  }{ (a-b)^{\alpha +2 } a^2} \right)~.
 \end{aligned}\end{equation}
 The final result, for complex $a,b$, reads
\begin{equation} \begin{aligned}\label{FTalpha}
    FT_\alpha&= 2^{\alpha } \sin \left(\frac{\pi  \alpha
   }{2}\right) \frac{\Gamma (\alpha
   +1)}{\pi^2}  \left( \frac{1}{b^{\alpha+1} a^2 }+\frac{1}{a^{\alpha+1} b^2 }  + \frac{b-a (\alpha +2)  }{ (a-b)^{\alpha +2 } a^2}+ \frac{a-b (\alpha +2)  }{ (b-a)^{\alpha +2 } b^2} \right)~.
 \end{aligned}\end{equation}

From the second line in \eqref{final-angle} we can also quantify the integral when $a=b$; this is the relevant limit for the two-point function $\langle {\cal O}_{\Phi^2}(x){\cal O}_\Psi(y)\rangle$. In this case we get
\begin{equation} \begin{aligned}\label{FTalpha2}
    FT^{(a=b)}_\alpha&= 2^{\alpha}
   \sin \left(\frac{\pi 
   \alpha }{2}\right)\frac{ \Gamma (\alpha
   +1)}{\pi ^2} \frac{2}{a^{\alpha+3}}~.
 \end{aligned}\end{equation}

\bibliographystyle{ytphys}
\bibliography{all}
\end{document}